\documentclass[aps,prd,preprintnumbers,superscriptaddress,nofootinbib]{revtex4}
\usepackage{hyperref}
\hypersetup{colorlinks=true,linkcolor=purple,anchorcolor=blue,citecolor=blue, filecolor=blue,urlcolor=red,bookmarksnumbered=true,
	pdfview=FitB
}
\usepackage{subcaption}
\usepackage{color}
\usepackage{xcolor}
\usepackage{ulem}
\usepackage{csquotes}
\colorlet{purple1}{blue!70!red}
\colorlet{darkred}{red!50!black}
\usepackage{graphicx}
\usepackage{psfrag}
\usepackage{color}
\usepackage{comment}
\usepackage{amssymb}
\usepackage{amsmath}
\usepackage{epstopdf}
\usepackage{epsf}

\newcommand{\Ta}{H_T}
\newcommand{\Tb}{\tilde{H}_T}
\newcommand{\Tc}{E_T}
\newcommand{\Td}{\tilde{E}_T}
\usepackage{natbib}

\newcommand{\nslash}{\kern 0.2 em n\kern -0.50em /}
\newcommand{\kslash}{\kern 0.2 em k\kern -0.45em /}
\newcommand{\lslash}{\kern 0.2 em l\kern -0.50em /}
\newcommand{\pslash}{\kern 0.2 em p\kern -0.50em /}
\newcommand{\Sslash}{\kern 0.2 em S\kern -0.50em /}
\newcommand{\Pslash}{\kern 0.2 em P\kern -0.50em /}
\newcommand{\Dslash}{\kern 0.2 em D\kern -0.65em /\kern 0.15em}

\newcommand{\be}{\begin{eqnarray}}
\newcommand{\ee}{\end{eqnarray}}

\newcommand{\bfp}{{\bf p}_{\perp}}

\newcommand{\uvec}[1]{\boldsymbol{#1}}

\begin{document}
	
	\title{Gluon generalized parton distributions of the proton at non-zero skewness  }
	\author{Dipankar~Chakrabarti}
	\email{dipankar@iitk.ac.in} 
	\affiliation{Department of Physics, Indian Institute of Technology Kanpur, Kanpur-208016, India}
 
    \author{Poonam~Choudhary}
    \email{poonamch@iitk.ac.in} 
    \affiliation{Department of Physics, Indian Institute of Technology Kanpur, Kanpur-208016, India}

	\author{Bheemsehan~Gurjar}
	\email{gbheem@iitk.ac.in} 
	\affiliation{Department of Physics, Indian Institute of Technology Kanpur, Kanpur-208016, India}

  \author{Tanmay Maji}
  \email{tanmayphy@nitkkr.ac.in}
  \affiliation{Department of Physics, National Institute of Technology Kurukshetra, Kurukshetra-136119, India}

  \author{Chandan Mondal}
  \email{mondal@impcas.ac.cn}
  \affiliation{Institute of Modern Physics, Chinese Academy of Sciences, Lanzhou 730000, China}
  \affiliation{School of Nuclear Science and Technology, University of Chinese Academy of Sciences, Beijing 100049, China}
	
 \author{Asmita~Mukherjee}
	\email{asmita@phy.iitb.ac.in} 
	\affiliation{Department of Physics, Indian Institute of Technology Bombay, Powai, Mumbai 400076, India}
	
%
\begin{abstract}
Using a recently developed light-front spectator model that incorporates gluon, where the light-front wave functions are {modeled} from the soft-wall AdS/QCD prediction, we examine the leading twist gluon generalized parton distributions (GPDs) inside the proton. We derive the chirally even and odd distributions by using the overlap representation of the {light-front} wave functions. In terms of GPDs at non-zero skewness, we investigate the entire three-dimensional representation of gluons. We analyse the gluon impact parameter distributions at $\xi=0$ using the Fourier transform of GPDs. We address the total angular momentum contribution of the gluons by using the Ji's sum rule and also give our predictions for both the canonical and kinetic orbital angular momentum  { in the} {light-cone gauge}. 
\end{abstract}
\maketitle
\section{Introduction}
One of the most difficult tasks in hadronic physics is to understand the three-dimensional structure of nucleons in terms of its constituent quarks and gluons, although parton distribution functions (PDFs) describe the distributions of the longitudinal momentum and polarization carried by quarks and gluons in a fast-moving hadron. The PDFs are the diagonal, or forward matrix elements of particular operators, providing a probability interpretation in terms of distributions and examining space-time correlations along the light-cone more extensively. It is only possible to fully understand the correlations by taking into account the non-diagonal, {or off-forward}  matrix elements of the same operators. These non-diagonal matrix elements can be parametrized in terms of generalized parton distributions (GPDs) \cite{Muller:1994ses,Diehl:2003ny,Belitsky:2005qn,Goeke:2001tz,Boffi:2007yc,Brodsky:2006in}. The GPDs encode spatial as well as partonic spin structure in a nucleon. {The GPDs can be related to the generalized transverse momentum distributions (GTMDs)~\cite{Meissner:2009ww,Lorce:2013pza,Kanazawa:2014nha,Mukherjee:2014nya,Hagiwara:2016kam,Ojha:2022fls,Maji:2022tog,Mukherjee:2015aja,More:2017zqp,Chakrabarti:2019wjx,Maji:2017ill,Zhang:2023xfe}. In the forward limit, or when the momentum transfer is zero, they reduce to  PDFs, that are accessible through inclusive processes. The second moment of the GPDs is related to the gravitational form factors (GFFs), which are the form factors of the energy-momentum tensor \cite{Ji:1996nm,Radyushkin:1997ki,Ji:1998xh,Ji:1998pc,Blumlein:1999sc,Goeke:2001tz,Guo:2023pqw,Hackett:2023nkr,GarciaMartin-Caro:2023klo,Burkert:2023wzr,More:2023pcy,Choudhary:2022den,deTeramond:2021lxc}. The GPDs or the GFFs are experimentally accessed through exclusive processes such as deeply virtual Compton scattering (DVCS) or deeply virtual vector-meson productions (DVMP). The GFFs, and therefore, the GPDs,} contain important information about the mass, angular momentum and mechanical properties of the nucleon~\cite{Polyakov:2018zvc, Lorce:2018egm,Chakrabarti:2020kdc,Mondal:2017wbf,Li:2023izn,Burkert:2023atx,Choudhary:2022den,Singh:2023hgu,More:2023pcy,More:2021stk}.
The GPDs are described as functions of three variables: longitudinal momentum fraction $x$ of the parton, longitudinal momentum fraction transferred in the process which is given by skewness $\xi$ and square of the total momentum transferred $t=\Delta^{2}$. 
In addition to providing information on the longitudinal behavior in momentum space along the direction in which the nucleon is moving, the GPDs also provide insights into how partons are spatially distributed in the transverse plane, making them an effective tool for studying hadron structure in three dimensions~\cite{Burkardt:2002hr,Burkardt:2005td}. Experimentally, determining GPDs in unique processes is quite difficult and demands high luminosity and resolution. First focused studies were therefore designed and carried out in the last decade \cite{H1:1999pji,HERMES:2001bob}. Through the {production} of vector mesons, they are experimentally accessible. {Data from experiments like H1, ZEUS, HERMES \cite{H1:2001nez,H1:2005gdw,H1:2009wnw,ZEUS:2003pwh, HERMES:2010nas, HERMES:2012gbh, HERMES:2012idp} have played a major role in the effort to extract the GPDs, also these will be explored in the JLab 12 GeV upgrade as well as the upcoming electron-ion collider (EIC) \cite{AbdulKhalek:2021gbh}.}  
The Fourier transform of GPDs with respect to the transverse momentum transfer, {$\Delta_\perp$}, gives the impact parameter dependent parton distributions, also known as  IPDs, which tell us how the partons of a given longitudinal momentum are distributed in transverse position space~\cite{Burkardt:2002hr,Burkardt:2005td}. Impact parameter measures the transverse separation of an active parton from the system's center of momentum. The spin densities can be represented in terms of impact parameter dependent GPDs for different proton polarizations~\cite{Lin:2023ezw,Maji:2017ill,Guo:2022upw,Mukherjee:2015aja,Mondal:2015uha}. {A more general description of the nucleon is given in terms of the GTMDs~\cite{Lorce:2011kd,Lorce:2011ni}, that help to access spin-spin and spin-orbit correlation of the gluon \cite{Bhattacharya:2022vvo}. The GTMDs are called mother distributions as the TMDs can be extracted from the GTMDs in the forward limit $\Delta_{\perp}=0$, and the GPDs can be obtained in impact parameter space upon integration over $\bfp$. 
The two-dimensional Fourier transform of GTMDs yields Wigner distributions~\cite{Ji:2003ak, Belitsky:2003nz}, 
which characterize the five-dimensional phase-space distribution of partons within the nucleon. These distributions play a crucial role in revealing spin-orbital correlations, essential for extracting parton orbital angular momentum~\cite{Ji:2012sj,Mukherjee:2015aja,Hatta:2011ku,More:2017zqp}.} 

{ Although it is known that gluon contributes significantly to the proton spin, the gluon distributions and contribution to the spin \cite{Jaffe:1995an} and OAM are less explored as compared to the quarks.  This is because, in most phenomenological models of the proton, there is no gluon. Only very recently, studies have been started in this direction \cite{PHENIX:2008swq,Boer:2011fh,Bhattacharya:2023jkd,Bhattacharya:2023yvo,Gurjar:2022jkx,Xu:2022abw,Chakrabarti:2023djs,Zhou:2022wzm,STAR:2021mqa,Ji:2020ena,Yang:2016plb,Sato:2016tuz,deFlorian:2014yva}. In this work, we use a recently developed spectator model \cite{Chakrabarti:2023djs} that incorporates the active gluon, to investigate the gluon GPDs at nonzero skewness ($\xi\neq 0$)} in the proton. There are eight gluon GPDs for the proton at leading-twist. Four of them ($H^g$, $E^g$, $\tilde{H}^g$, and $\tilde{E}^g $) are chirally even  and other four GPDs ($H_T^g$, $E_T^g$ $\tilde{H}_T^g$, and  $\tilde{E}_T^g $) are chirally odd. {The GPDs $H^g$ and $\tilde{H}^g$ are related to the unpolarized and polarized gluon PDFs, respectively. In the impact parameter space, they provide the spin densities of the unpolarized and circularly polarized gluons} in a longitudinally polarized proton. The  GPD $E^g$ is related to the distortion of the unpolarized gluon distribution in a transversely polarized proton,  in some models, this GPD is related to the gluon Sivers function which is a T-odd TMDPDF~\cite{Meissner:2007rx}. Chiral odd GPDs are also accessible in specific processes like deeply virtual Compton scattering or deeply virtual meson generation. They are more difficult to probe experimentally than chiral even GPDs, as they need another chiral-odd object in the amplitude to combine.

{As stated above, the GTMDs as well as the GPDs are important to understand the angular momentum contribution of the gluons to the nucleon spin, and its orbital angular momentum (OAM). The decomposition of the total angular momentum of the nucleon into spin and OAM is not unique \cite{Leader:2013jra}. There are two well-known decompositions, known as the kinetic and canonical decompositions, available in the literature~\cite{Wakamatsu:2010qj,Ji:1996ek,Leader:2013jra}.
We also present a calculation  of the gluon contribution to the total angular momentum of the proton as well as  the OAM in different decompositions in our model.}

The work is arranged in the following manner: In section \ref{sec:model}, we briefly review the light front gluon spectator model. In section \ref{sec:GPDs intro}, we discuss the general definition of gluon GPD correlator which allows us to assess both chiral-even and chiral-odd GPDs at leading twist with both zero and non-zero skewness. In section~\ref{OAM} we evaluate the chiral-even gluon GTMDs and we present a discussion on gluon orbital angular momentum and spin-correlation distributions.
In section~\ref{sec:IPDs} we present the impact parameter dependent generalized parton distributions at zero skewness. Finally, in Section~\ref{sec:conclusion} we present our conclusions and outlook. Many supplementary formulae explaining the details of the parametrization are presented in the Appendix.
%
%
\section{light front wave function }\label{sec:model}
 To investigate the {gluon distribution functions within the proton}, {we have recently formulated} a light-front spectator model for the proton in which gluon is considered as an active {parton}. A detailed description about the model construction can be found in Ref.~\cite{Chakrabarti:2023djs}. Apart from the struck gluon, the remaining components, including the three valence quarks and any additional gluons or sea quarks, collectively form the spectator system. We consider the proton as a {composite} system of spin-1 gluon and spin-1/2 spectator. \\
 
The model is based on the light-cone approach with $x^{\pm} = x^0 \pm x^3 $. We choose a reference frame where the transverse momentum of the proton vanishes, i.e., $ P=(P^+,\frac{M^2}{P^+},\textbf{0}_\perp)$. The momentum of active parton  is given by $ p=(x P^+, \frac{p^2+\textbf{p}_\perp^2}{x P^+}, \textbf{p}_\perp)$ and the momentum of the spectator $ P_X=((1-x) P^+, P^-_X, -\textbf{p}_\perp)$. The variable $x=p^{+}/P^{+}$ represents the fraction of longitudinal momentum carried by the struck gluon. Therefore, the proton state can be written as a two-particle Fock-state expansion with proton helicity $J_{z}=\pm\frac{1}{2}$ as~\cite{Brodsky:2000ii},
 \begin{eqnarray}\label{state}\nonumber
		|P;\uparrow(\downarrow)\rangle
		= \int \frac{\mathrm{d}^2 \bfp \mathrm{d} x}{16 \pi^3 \sqrt{x(1-x)}}\times \Bigg[\psi_{+1+\frac{1}{2}}^{\uparrow(\downarrow)}\left(x, \bfp\right)\left|+1,+\frac{1}{2} ; x P^{+}, \bfp\right\rangle+\psi_{+1-\frac{1}{2}}^{\uparrow(\downarrow)}\left(x, \bfp \right)\left|+1,-\frac{1}{2} ; x P^{+}, \bfp \right\rangle\\ 
		+\psi_{-1+\frac{1}{2}}^{\uparrow(\downarrow)}\left(x, \bfp \right)\left|-1,+\frac{1}{2} ; x P^{+}, \bfp \right\rangle+\psi_{-1-\frac{1}{2}}^{\uparrow(\downarrow)}\left(x, \bfp\right)\left|-1,-\frac{1}{2} ; x P^{+}, \bfp\right\rangle\bigg],
	\end{eqnarray}
	where $\psi_{\lambda_{g}\lambda_{X}}^{\uparrow(\downarrow)}(x,\bfp)$ are the two-particle LFWFs, which represents the probability amplitudes to find constituents corresponding to the two-particle Fock state $|\lambda_{g},\lambda_{X};xP^{+},\bfp \rangle$ with longitudinal momentum $xP^{+}$, transverse momentum ${\bf p}_{\perp}$ and helicities $\lambda_{g}$ and $\lambda_{X}$ in the proton. Here $\lambda_{g}$ and $\lambda_{X}$ {stand} for the helicity components of the active gluon and spectator, respectively.

 {The two-particle {LFWFs} of the proton with $J_{z}=+1/2$ have the following form }~\cite{Chakrabarti:2023djs},
	\begin{eqnarray} \label{LFWFsuparrow}   \nonumber
		\psi_{+1+\frac{1}{2}}^{\uparrow}\left(x,\bfp\right)&=&-\sqrt{2}\frac{(-p^{1}_{\perp}+ip^{2}_{\perp})}{x(1-x)}\varphi(x,\bfp^2), \\ \nonumber
		\psi_{+1-\frac{1}{2}}^{\uparrow}\left(x, \bfp\right)&=&-\sqrt{2}\bigg( M-\frac{M_{X}}{(1-x)} \bigg) \varphi(x,\bfp^2), \\ \nonumber
		\psi_{-1+\frac{1}{2}}^{\uparrow}\left(x, \bfp\right)&=&-\sqrt{2}\frac{(p^{1}_{\perp}+ip^{2}_{\perp})}{x}\varphi(x,\bfp^2), \\
		\psi_{-1-\frac{1}{2}}^{\uparrow}\left(x, \bfp\right)&=&0,
	\end{eqnarray}
	where both $M$ and $M_X$ represent the masses of the proton and spectator. {Similarly, the two-particle {LFWFs}  of a proton with $J_{z}=-1/2$ have  the form},
	\begin{eqnarray} \label{LFWFsdownarrow}   \nonumber
		\psi_{+1+\frac{1}{2}}^{\downarrow}\left(x, \bfp\right)&=& 0, \\ \nonumber
		\psi_{+1-\frac{1}{2}}^{\downarrow}\left(x,\bfp\right)&=&-\sqrt{2}\frac{(-p^{1}_{\perp}+ip^{2}_{\perp})}{x}\varphi(x,\bfp^2), \\ \nonumber
		\psi_{-1+\frac{1}{2}}^{\downarrow}\left(x, \bfp\right)&=&-\sqrt{2}\bigg( M-\frac{M_{X}}{(1-x)} \bigg) \varphi(x,\bfp^2),  \\
		\psi_{-1-\frac{1}{2}}^{\downarrow}\left(x, \bfp \right)&=& -\sqrt{2}\frac{(p^{1}_{\perp}+ip^{2}_{\perp})}{x(1-x)}\varphi(x,\bfp^2).
	\end{eqnarray}
  Where $\varphi(x,\bfp^2)$ is the modified form of the soft-wall AdS/QCD wave function, which is modeled by introducing the parameters $a$ and $b$.
The expression for the wave function $\varphi(x,\bfp^2)$ is 
modified by $x$ and $1-x$ factors, which explains the small and large $x$ region behaviour {of the PDFs}. The complete form of the modified soft-wall AdS/QCD wave function is given as~\cite{Gutsche:2013zia,Chakrabarti:2023djs},
\begin{eqnarray}\label{AdSphi}
\varphi(x,\bfp^2)=N_{g}\frac{4\pi}{\kappa}\sqrt{\frac{\log[1/(1-x)]}{x}}x^{b}(1-x)^{a}\exp{\bigg[-\frac{\log[1/(1-x)]}{2\kappa^{2}x^2}\bfp^{2}\bigg]}
\end{eqnarray}
where {$\kappa$ is the AdS scale parameter which is given as $\kappa=0.4$ GeV~\cite{Brodsky:2014yha}}. While, $a,~b$, and $N_{g}$ are the model parameters which are fixed by fitting the NNPDF3.0 NLO gluon unpolarized PDF data set at initial scale $\mu_0=2$ GeV and can be found in Ref.~\cite{Chakrabarti:2023djs}. For the stability of the proton, the sum of constituent masses is considered higher than the proton mass, i.e., $M_{X}>M$ \cite{Bacchetta:2020vty}.
%
%
\section{Gluon GPDs} \label{sec:GPDs intro}

In the light-cone gauge $A^{+}=0$, the off-forward matrix elements of the bilocal currents of light-front correlation functions define the {four leading twist} gluon helicity conserving GPDs~\cite{Diehl:2001pm,Diehl:2003ny,Ji:1998pc}
\begin{align}
 \frac{1}{P^+} \int \frac{d z^-}{2\pi}\, e^{ix P^+ z^-}
  \langle p',\lambda'|\, 
     F^{+i}(-{\textstyle\frac{z}{2}})\, 
     F^{+i}({\textstyle\frac{z}{2}})\, 
  \,|p,\lambda \rangle \Big{|}_{\substack{z^+=0\\\mathbf{z}_{T}=0}}= \frac{1}{2P^+} \bar{u}(p',\lambda')
  \left[
  H^g\, \gamma^+ +
  E^g\, \frac{i \sigma^{+\alpha} \Delta_\alpha}{2M} 
  \right] u(p,\lambda) ,\\
   - \frac{i}{P^+} \int \frac{d z^-}{2\pi}\, e^{ix P^+ z^-}
  \langle p',\lambda'|\, 
     F^{+i}(-{\textstyle\frac{z}{2}})\, 
          \tilde{F}^{+i}({\textstyle\frac{z}{2}})\, 
  \,|p,\lambda \rangle \Big{|}_{\substack{z^+=0\\\mathbf{z}_{T}=0}}
= \frac{1}{2P^+} \bar{u}(p',\lambda') \left[
  \tilde{H}^g\, \gamma^+ \gamma_5 +
  \tilde{E}^g\, \frac{\gamma_5 \Delta^+}{2M}\, \, 
  \right] u(p,\lambda) ,
\end{align}
{where $\tilde{F}^{\alpha\beta}=\frac{1}{2}\epsilon^{\alpha\beta\gamma\delta}F_{\gamma\delta}$ is the dual field strength tensor and a summation over $i=1,2$ is implied.} 
Similarly, the {remaining four} gluon helicity flip GPDs involves the matrix elements of {gluon tensor operator} $\mathbf{S}F^{+i}(-z/2)F^{+j}(z/2)$, {where {\bf S} represents the symmetrization operator in $i$ and $j$} and can be given as,
\begin{align}
  \label{flip-gluon}
- \frac{1}{P^+} \int \frac{d z^-}{2\pi}\, e^{ix P^+ z^-}&
  \langle p',\lambda'|\, {\mathbf S}
     F^{+i}(-{\textstyle\frac{z}{2}})\,
F^{+j}({\textstyle\frac{z}{2}})
  \,|p,\lambda \rangle \Big{|}_{\substack{z^+=0\\\mathbf{z}_{T}=0}} 
={\mathbf S}\,
\frac{1}{2 P^+}\, \frac{P^+ \Delta^j - \Delta^+ P^j}{2 M P^+}\nonumber \\
&
\times \bar{u}(p',\lambda') \Bigg[
  \Ta^g\, i \sigma^{+i}+\Tb^g\, \frac{P^+ \Delta^i - \Delta^+ P^i}{M^2} 
  +\Tc^g\, \frac{\gamma^+ \Delta^i - \Delta^+ \gamma^i}{2M} +
  \Td^g\, \frac{\gamma^+ P^i - P^+ \gamma^i}{M}\, 
   \Bigg] u(p,\lambda),
\hspace{2em} 
\end{align} 
{where $u$ $\left(\bar{u}\right)$ are the light-front spinors with $p$ $(p')$ and $\lambda$ $(\lambda')$ the momenta and the helicity of the initial (final) state of proton, respectively.  
In the symmetric frame, the kinematic variables are denoted as: the average momentum $P^\mu=\frac{(p+p')^\mu}{2}$, momentum transfer $\Delta^\mu=p'^\mu-p^\mu$, the skewness $\xi=-\Delta^+/2P^+$, and the invariant momentum transfer in the process $t=\Delta^2$.}
 The standard form of the gluon field strength tensor $F_a^{\mu\nu}(x)$ is given in terms of structure constant  $f_{abc}$ as follows
\begin{align}
F^{\mu\nu}_a(x)=\partial^\mu A^\nu_a(x)-\partial^\nu A_a^\mu(x)+gf_{abc}A_b^\mu(x)A_c^\nu(x).
\end{align}
{{We choose light cone gauge $A^+=0$, which fixes the gauge link between the fields to be unity. It also implies that $F^{+i} = \partial^+ A^i$ which simplifies the calculation. The operator $F^{+i}(-\frac{z}{2})$ is associated with incoming gluon while $F^{+j}(\frac{z}{2})$ is associated with an
outgoing gluon. We use the light-cone quantization framework to obtain the gluon distributions in terms of light-cone helicity amplitudes. We then define the following helicity amplitudes  which are connected to different gluon GPDs based on the different proton and gluon helicity configurations~\cite{Diehl:2003ny,Diehl:2001pm,Boffi:2007yc,Maji:2017ill}:}}
%
\begin{eqnarray}
  \label{helicity-elements-gluon}
A_{\lambda'\mu', \lambda\mu} =
\frac{1}{P^{+}} \int \frac{d z^{-}}{2\pi}\, e^{ix P^{+} z^{-}} \langle p',\lambda'|\, 
     \epsilon^{i}(\mu')F^{+i}(-{\textstyle\frac{z}{2}})\, 
     F^{+j}({\textstyle\frac{z}{2}})\epsilon^{*j}(\mu) \,|p,\lambda 
   \rangle\Big|_{z^+=0,\,\mathbf{z}_T=0} \, 
\label{helicity amp}
\end{eqnarray}
where $\mu$ $(\mu')$ denotes the the gluon helicity of the initial (final) state {and $\epsilon$ is the two-dimensional gluon polarization vectors}. Parity invariance results following relation among the various helicity amplitudes as,
\begin{align}\label{parity}
A_{-\lambda'-\mu', -\lambda-\mu} = (-1)^{\lambda'-\mu'-\lambda+\mu}\,(A_{\lambda'\mu', \lambda\mu})^{\ast}
\end{align}

\subsection{GPDs at non zero skewness}	\label{sec:GPDs at non zero skewness}


The internal structure of the proton has primarily been investigated within the confines of the $\xi=0$ limit in most studies. However, in order to fully uncover the wealth of information available within the three-dimensional momentum transfer space, it is imperative to assess the 
GPDs at non-zero skewness values. Recent theoretical works \cite{Guo:2023ahv,Guo:2023qgu} have started to explore GPDs at non-zero skewness. {In fact, skewness is typically non-zero in experiments. The GPDs evolve with the scale differently in different kinematical regions~\cite{Diehl:2000xz}: for the quark GPDs, (i)  $x \geq \xi $ corresponds to the situation when an active quark of initial longitudinal momentum fraction of $ x+\xi$ struck with a photon and come back to the nucleon with longitudinal momentum fraction $x-\xi$; (ii)  $x \leq -\xi $ corresponds the distributions of anti-quarks where both the longitudinal momentum fractions  $ x+\xi$ and  $ x-\xi$ are negative; both the  $x \geq \xi $  and $x \leq -\xi $ regions are known as Dokshitzer-Gribov-Lipatov-Altarelli-Parisi (DGLAP) domain; (iii) in the third region $-\xi \leq x \leq \xi$, which is known as Efremov-Radyushkin-Brodsky-Lepage (ERBL) region the GPDs correspond to the process in which a quark of longitudinal momentum $ x+\xi$ interact with a virtual photon and emit an anti-quark of longitudinal momentum $\xi-x$. Here, we investigate the gluon GPDS, in which the active parton is a gluon,  in the DGLAP region ($x>\xi$ only), as in the ERBL region, the GPDs would involve particle number changing overlaps involving higher Fock components of the LFWFs, which is beyond the scope of the present work.}  
{The helicity amplitudes}, $A_{\lambda'\mu', \lambda\mu} $ include specific information regarding the initial and final helicities of the proton and gluon. The chiral-even GPDs are defined in terms of helicity amplitudes wherein the helicity of the gluon does not change and the helicity of the nucleon does. While in the chiral odd GPDs, gluon flips its helicity. 

In the reference frame where the momenta $\vec{p}$ and $\vec{p}^{\prime}$ lie in the $x-z$ plane, the chiral-even GPDs can be expressed through the gluon helicity conserving amplitudes within the helicity basis as follows~\cite{Diehl:2001pm,Maji:2017ill,Boffi:2007yc} : 
\be
H^g&=&\frac{1}{\sqrt{1-\xi^2}}T^g_1 - \frac{2M\xi^2} {\sqrt{t_0 - t}(1 - \xi^2)}T^g_3,\label{H}\\
E^g&=&-\frac{2M} {\epsilon \sqrt{t_0 - t}}T^g_3,\label{E}\\
\widetilde{H}^g&=&\frac{1}{\sqrt{1-\xi^2}}T^g_2+\frac{2M\xi} {\sqrt{t_0 - t}(1 - \xi^2)}T^g_4,\label{Ht}\\
\widetilde{E}^g&=&\frac{2M} {\epsilon \xi\sqrt{t_0 - t}}T^g_4,\label{Et}
\ee
while, the chiral-odd GPDs can be written in terms of gluon helicity non-conserving amplitudes as, 
\be
\Ta^g &=&\frac{2 M}{\epsilon  \sqrt{t_0-t}(1-\xi^2)}\widetilde{T}^g_{1} - \frac{4 M^2 \xi}{ (t_0-t)(1-\xi^2)\sqrt{1-\xi^2}}\widetilde{T}^g_{3} ,\\
\Tc^g&=&\frac{4 M^2}{ (t_0-t)(1-\xi^2)\sqrt{1-\xi^2}}\left( \xi \widetilde{T}^g_{3}+ \widetilde{T}^g_{4} \right) \\
\Td^g&=&\frac{4 M^2}{ (t_0-t)(1-\xi^2)\sqrt{1-\xi^2}}\left( \widetilde{T}^g_{3}+ \xi \widetilde{T}^g_{4} \right) \\
\Tb^g&=&0
\ee
where $T^g_{i}$ and $\widetilde{T}^g_{i}$  are chiral even and chiral-odd helicity basis which are defined as a combination of helicity amplitudes \cite{Kriesten:2021sqc} as: 
\be
T^g_{1} =A_{++,++} + A_{-+,-+}, &\quad&
T^g_{2} =A_{++,++} - A_{-+,-+},\nonumber\\
T^g_{3} =A_{++,-+} + A_{+-,--}. &\quad&
T^g_{4} =A_{++,-+} - A_{+-,--},
\ee
and \be
\widetilde{T}^g_{1} =A_{++,--} + A_{-+,+-}, &\quad&
\widetilde{T}^g_{2} =A_{++,--} - A_{-+,+-},\nonumber\\
\widetilde{T}^g_{3} = A_{++,+-}+A_{+-,++}, &\quad&
\widetilde{T}^g_{4} = A_{++,+-}- A_{+-,++}
\ee
\begin{figure}
	\centering
	\includegraphics[scale=0.29]{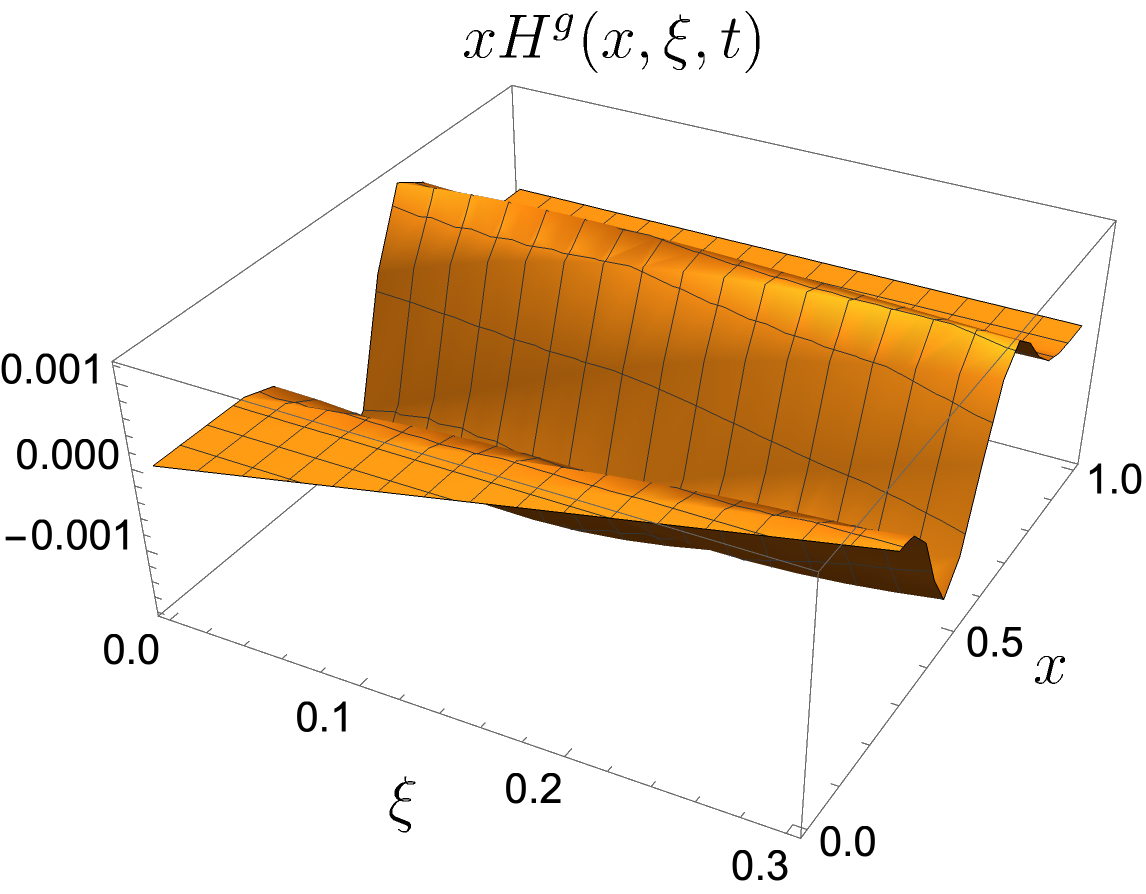}\hspace{0.2cm}
	\includegraphics[scale=0.3]{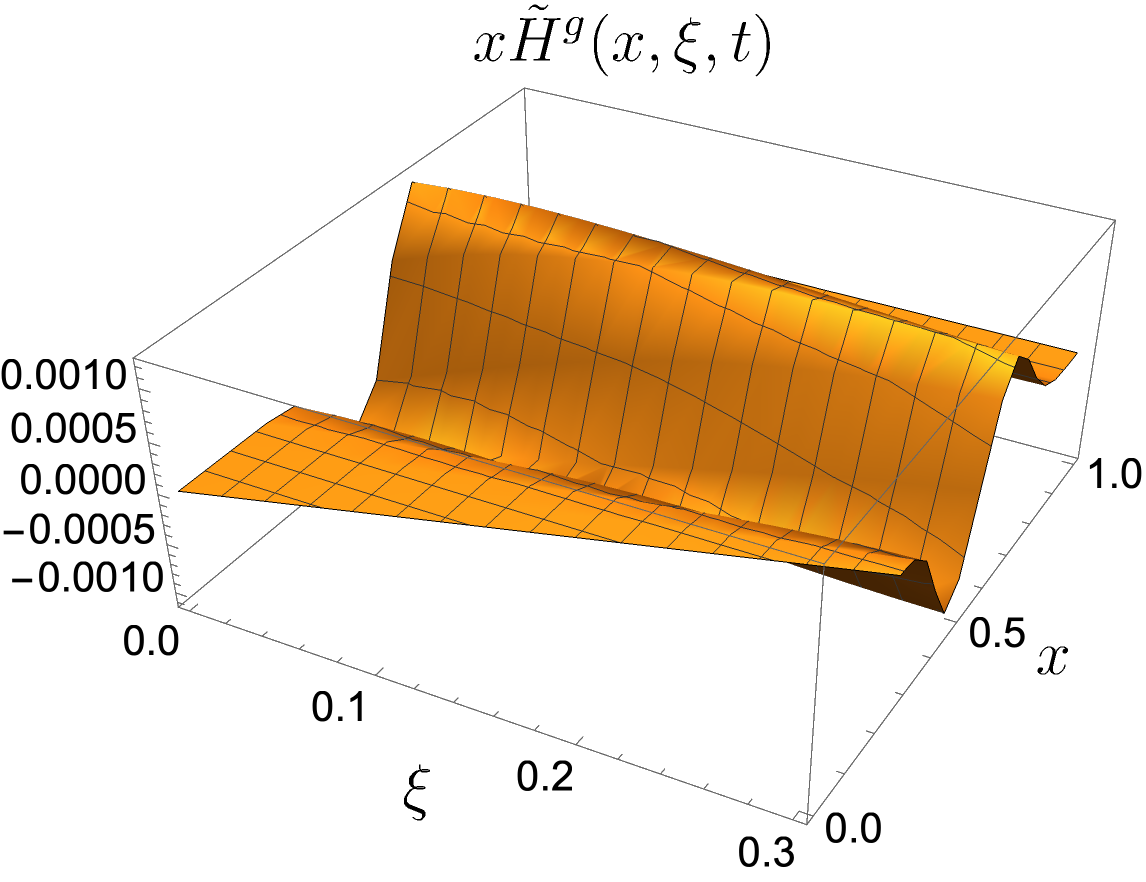}\hspace{0.2cm}
    \includegraphics[scale=0.3]{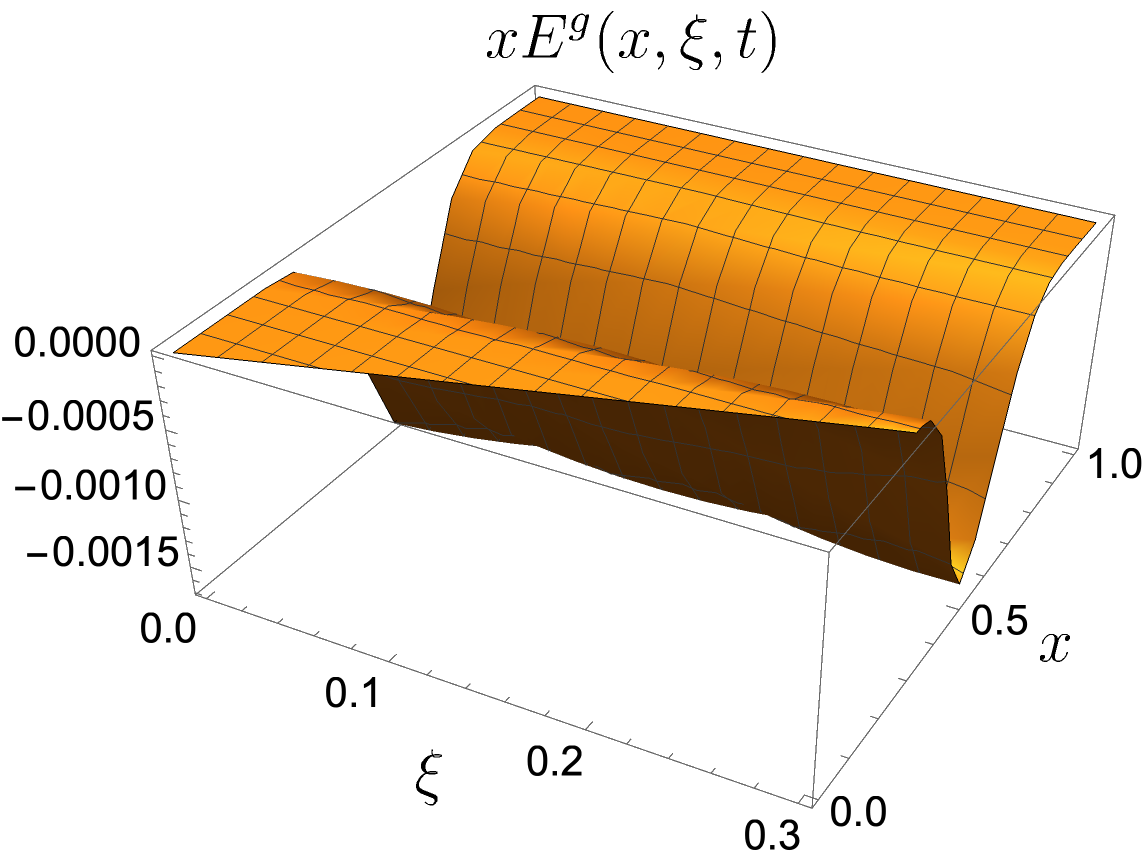}
    \vspace{0.2cm}
	\includegraphics[scale=0.29]{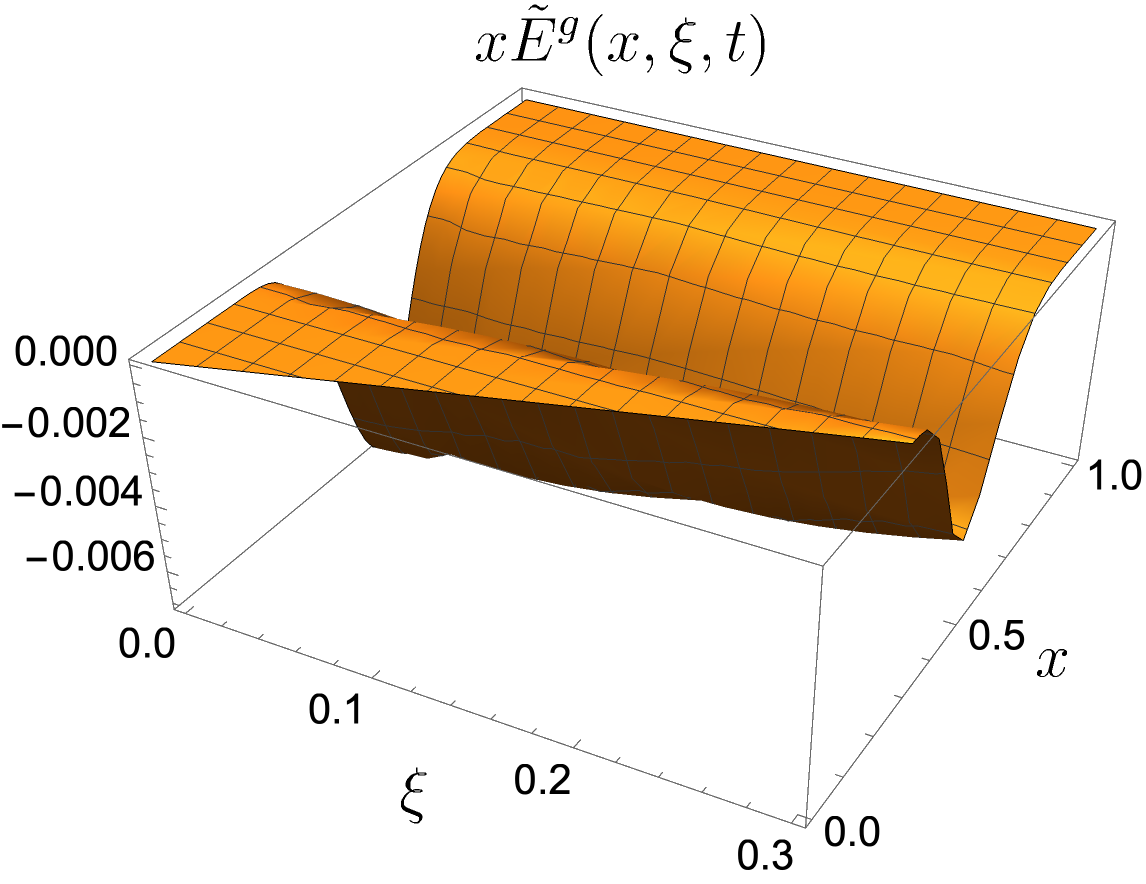}\hspace{0.2cm}
    \includegraphics[scale=0.3]{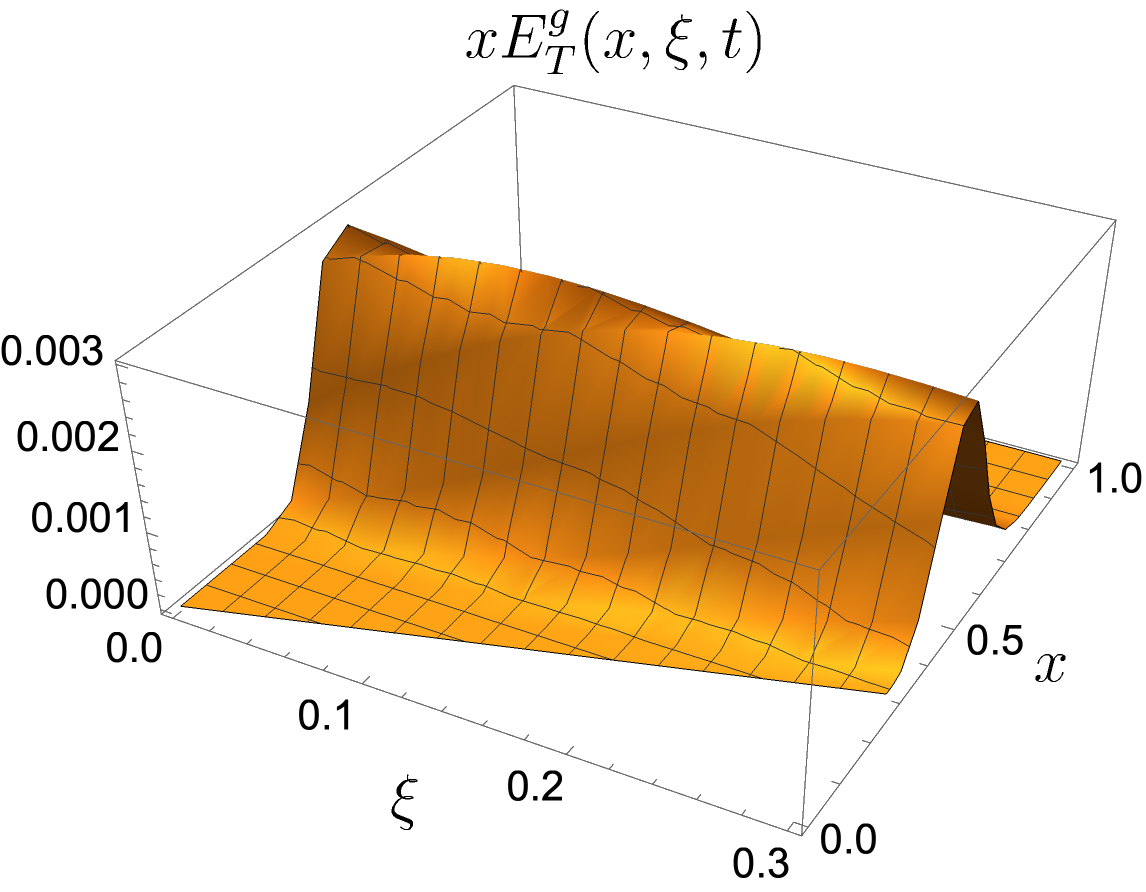} \hspace{0.2cm}
     \includegraphics[scale=0.29]{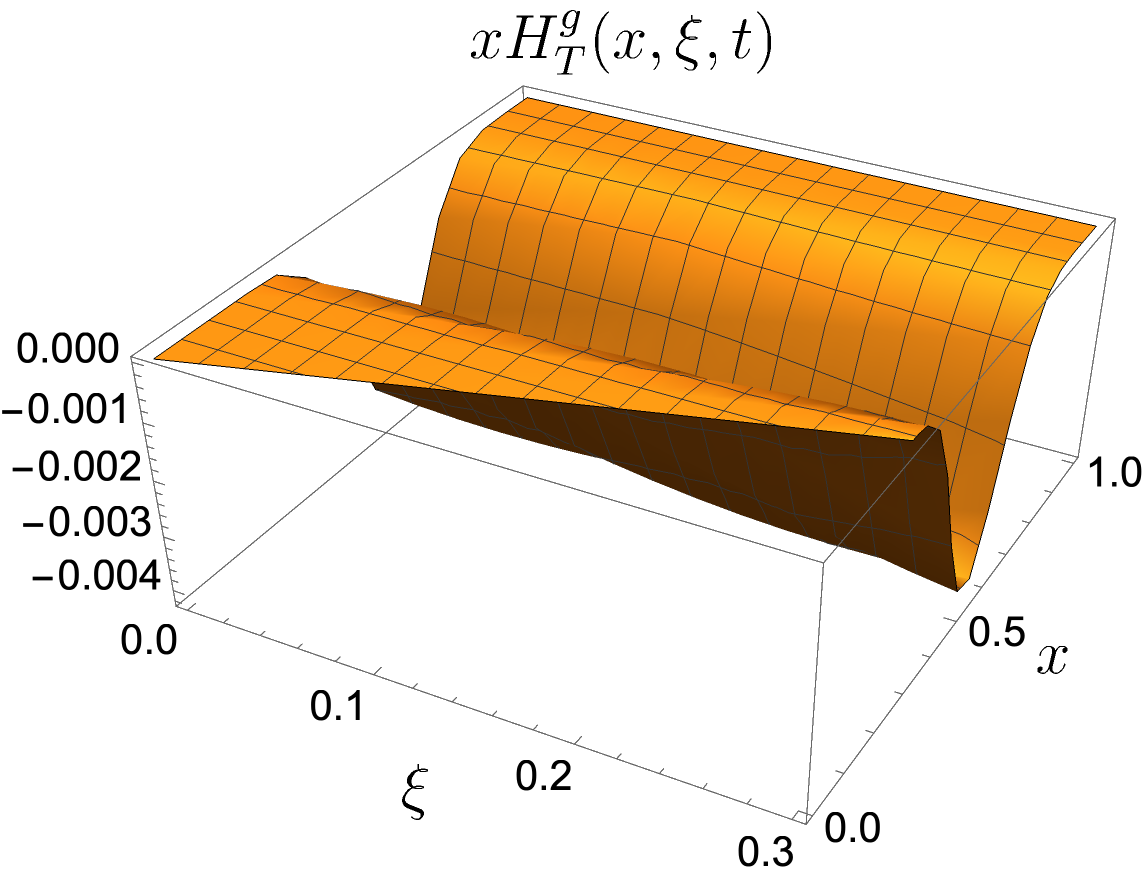} 
 \caption{
 3D representation of gluon GPDs as a function of gluon longitudinal momentum fraction $x$ and skewness parameter $\xi$ for fixed transverse momentum transfer $-\lvert t \rvert=3$ GeV$^2$.}
 \label{fig:3Dnonzeroskewness}
\end{figure}
\begin{figure}
	\centering
	\includegraphics[scale=0.28]{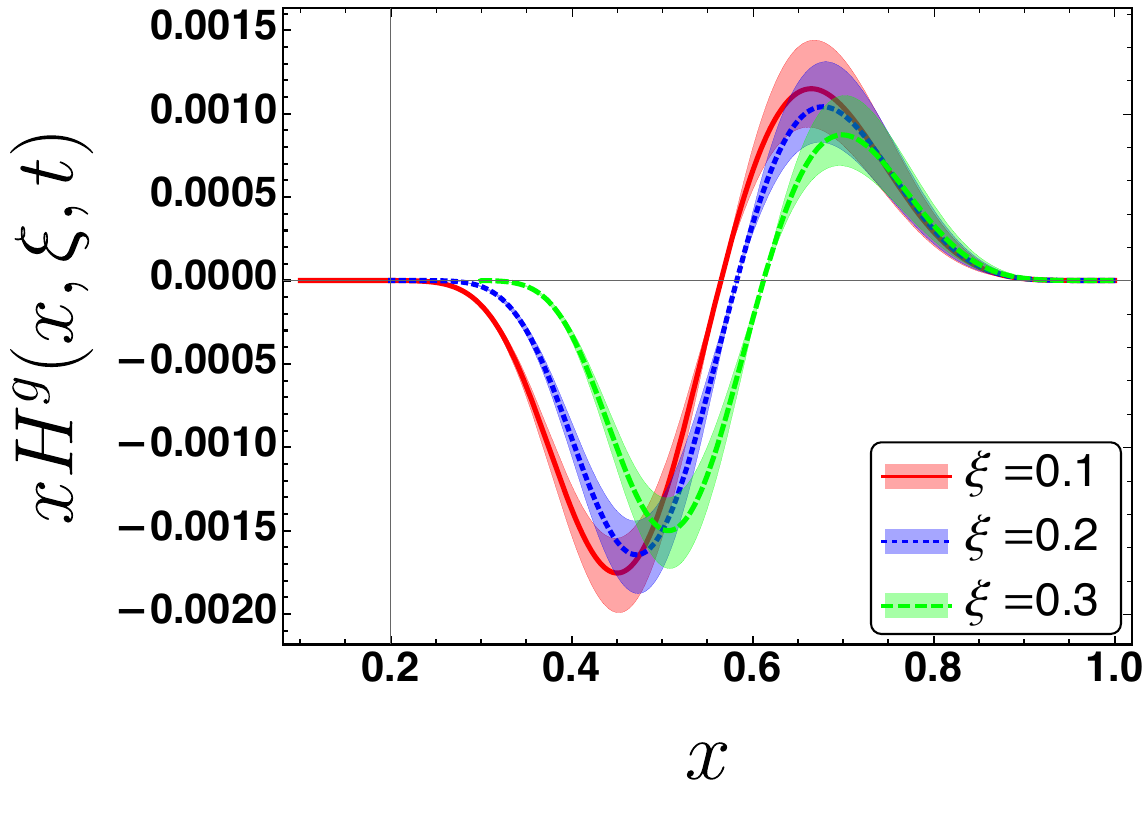}\hspace{0.2cm}
	\includegraphics[scale=0.28]{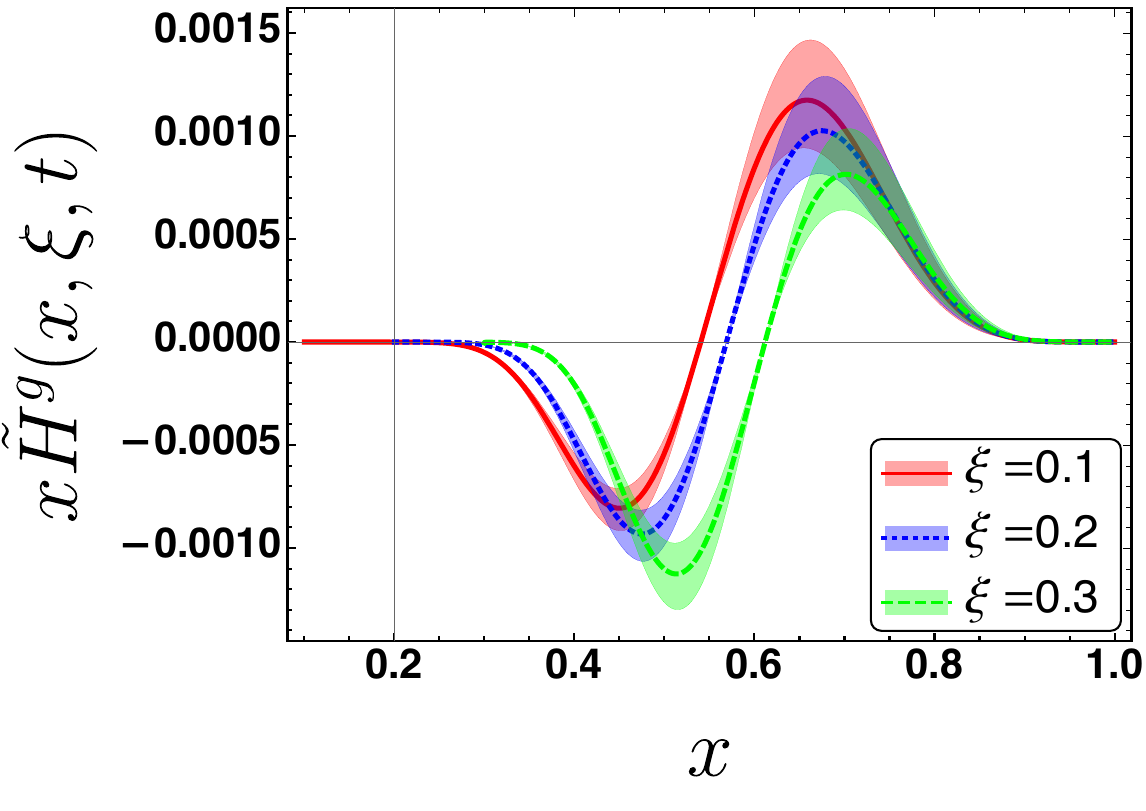}\hspace{0.2cm}
    \includegraphics[scale=0.28]{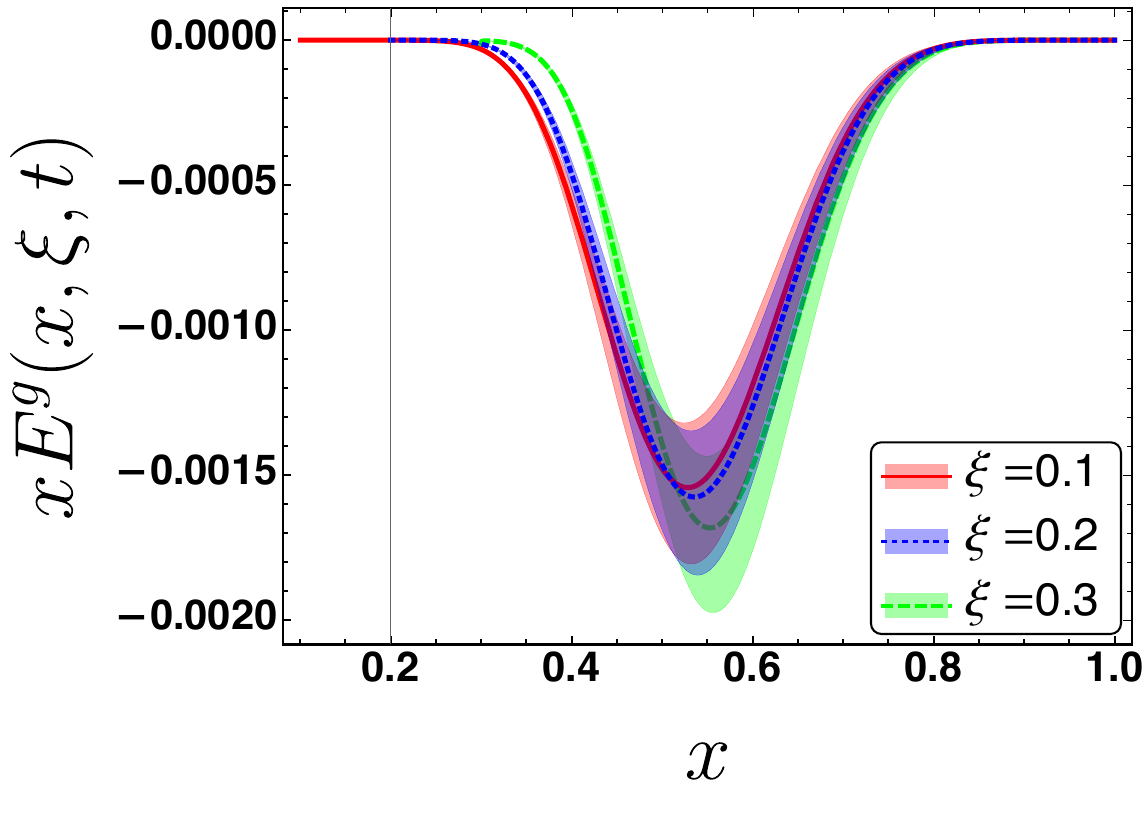}
    \vspace{0.2cm}
	\includegraphics[scale=0.28]{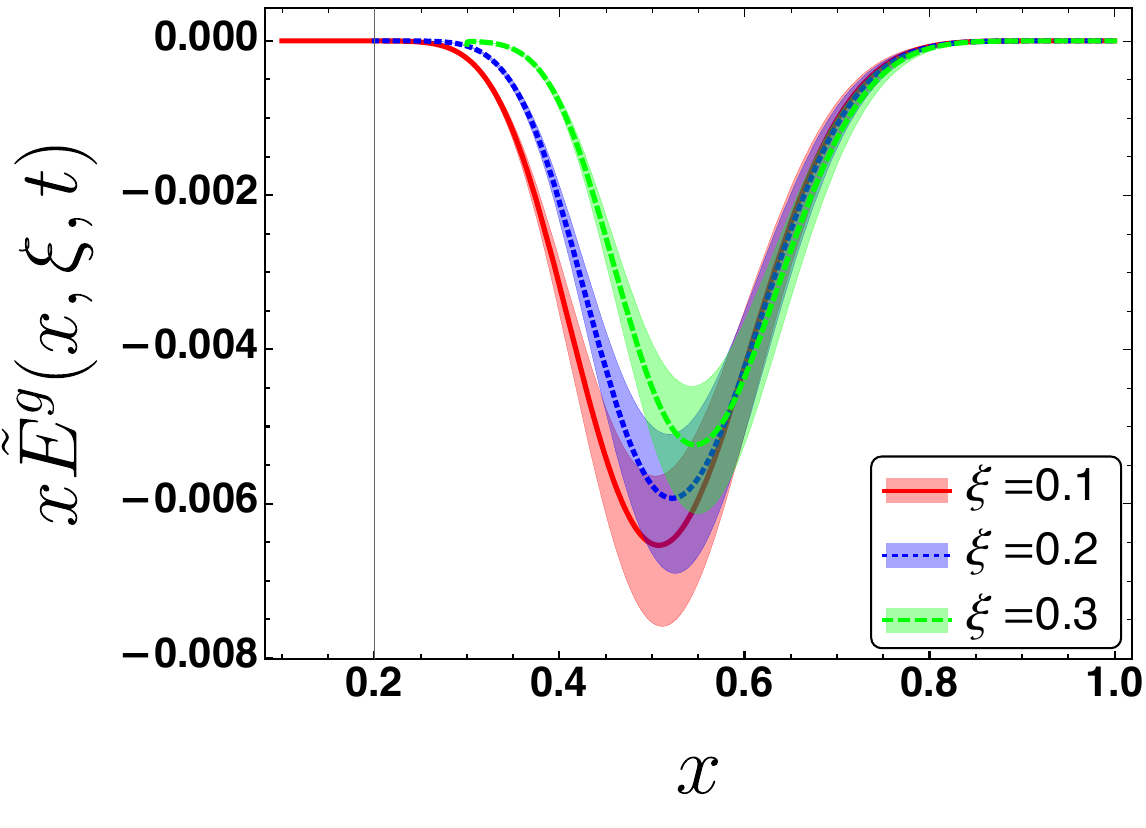}\hspace{0.2cm}
    \includegraphics[scale=0.28]{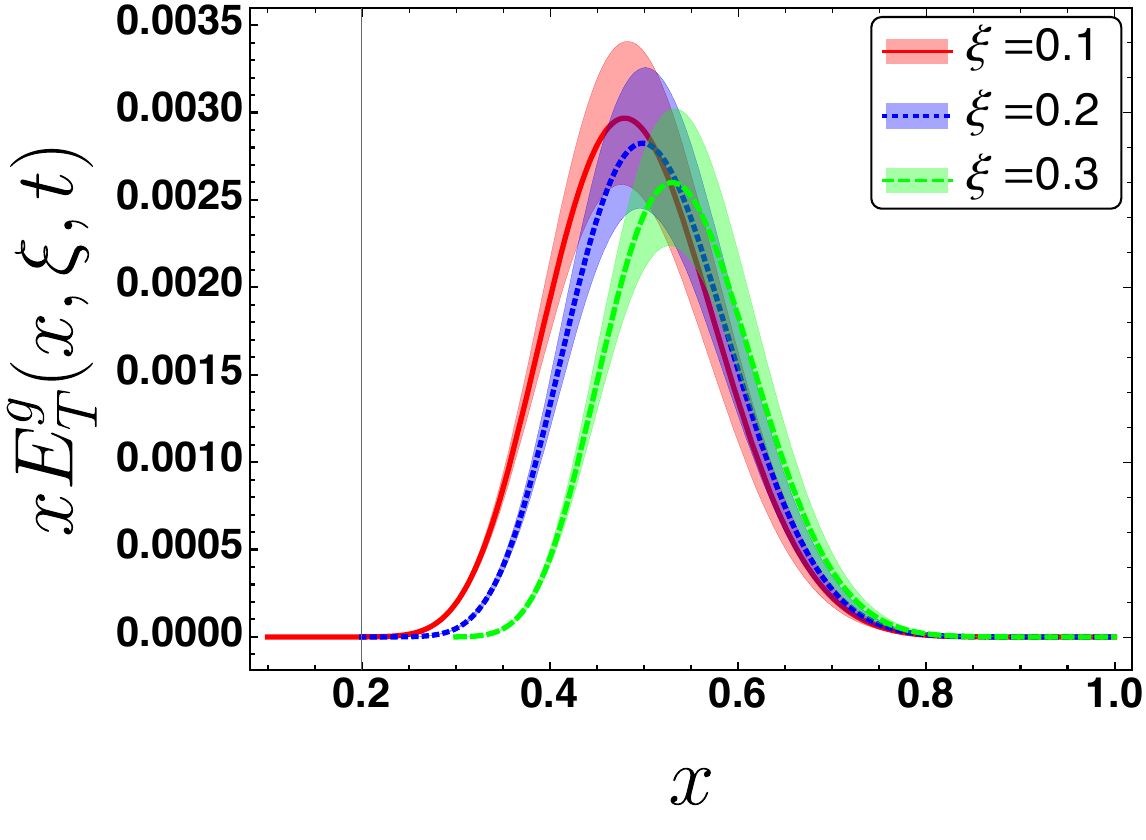} \hspace{0.2cm}
     \includegraphics[scale=0.28]{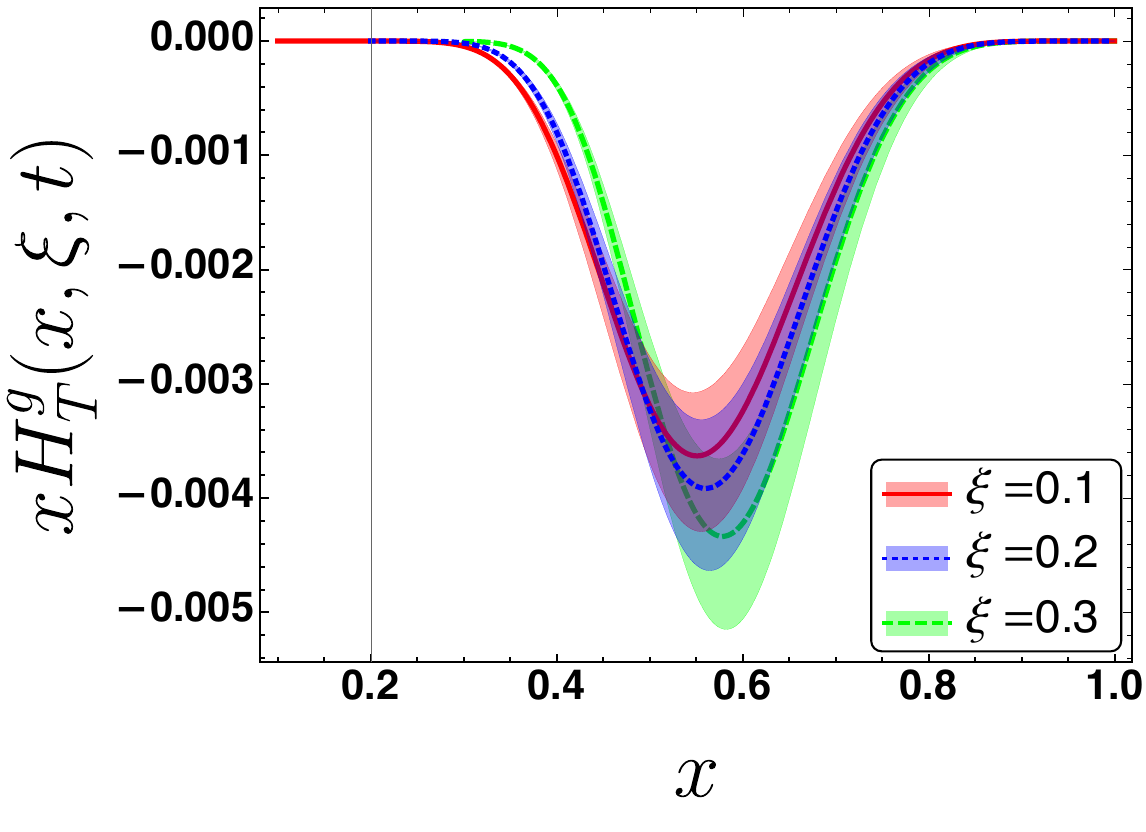}
 \caption{
 2D plots of gluon GPDs at $\xi=0.1,0.2$ and $\xi=0.3$ as a function of longitudinal momentum fraction $x$ for a fixed momentum transfer $-\lvert t \rvert=3 $ GeV$^2$.}
 \label{fig:DGLAPregion}
\end{figure}
Here $t_{0}$ is the minimum value of the transverse momentum transfer $t$, i.e., $t_{0}=-4M^{2}\xi^{2}/(1-\xi^{2})$ for a given value of $\xi$ with $\epsilon=\text{sgn}(D^{i})$ where $D^{i}$ is the $i^{th}$ component of $D^{\alpha}=P^{+}\Delta^{\alpha}-\Delta^{+}P^{\alpha}$ and $t=t_{0}$ for $D^{i}=0$.  The matrix elements $T_{i}^{g}$ and $\widetilde{T}_{i}^{g}$ can be written in terms of overlap of light-front wavefunctions. The detailed calculation and expressions for the above matrix elements are given in the Appendix \ref{append}. Using the equations (\ref{t1g} -\ref{t4tilde}) in the above relations we can get the expressions for individual GPDs at non-zero skewness. 

  Fig.~\ref{fig:3Dnonzeroskewness} depicts both the chiral even and chiral-odd GPDs in three-dimensional momentum space, with skewness parameter, $\xi$ and $x$ at a fixed transverse momentum transfer of $-\lvert t \rvert=3$ GeV$^2$. {Though the GPDs are oscillatory in $x$-space, they vary monotonically with $\xi$.} 
 As shown in Fig.~\ref{fig:3Dnonzeroskewness} the distributions exhibit non-zero values in the $x \geq \xi$ region only. The 3D distribution of  $x H^g$ shows the behavior of unpolarized GPD. 
 The amplitude of the distribution depends on the value of $x$ and $\xi$ in such a way that it is close to zero when $x$ is very close to $\xi$ but {the position of the peak in $x$ depends on the value of $\xi$}. A similar kind of behavior is also shown by $x \tilde{H}^g$ distribution with a magnitude almost half of the magnitude of the $x H^g$ distribution. The forward limit of $ H^g$ and $ \tilde{H}^g $ {can} be related to the unpolarized and helicity PDFs, respectively. Since the results in Fig.~\ref{fig:3Dnonzeroskewness} are presented at fixed momentum transfer, therefore we cannot relate these results to the forward limit. The distributions $x E^g$, $x \tilde{E}^g$, and $x H_{T}^g$ show similar behavior but {of} different orders of amplitude.

In Fig.~\ref{fig:DGLAPregion}, we show the representation of GPDs as a function of the longitudinal momentum fraction variable $x$ within the context of the DGLAP region. These plots are generated at specific values of skewness parameter $\xi$ and momentum transfer $-\lvert t \rvert=3$ GeV$^2$. The 2D distributions of $ x H^g$ and $ x \tilde{H}^g$ {oscillates from negative to positive} as they go from small to large $x$ for a particular value of $\xi$ {whereas $xE^g,~x\tilde{E}^g,~xE_T^g$ and  $xH_T^g$ don't change their signs.} The magnitude of both positive and negative peaks depends on the value of $\xi$ as shown in Fig.~\ref{fig:DGLAPregion}. We observe that the magnitudes of the peaks {$ x H^g$ and $ x \tilde{H}^g$} decrease and shift towards the large value of $x$ as we increase the value of the skewness parameter. The distributions $x \tilde{E}^g$ and $x E_{T}^g$ follow a similar trend. {While, contrary to these, the magnitudes of} $x E^g$ and $x H_{T}^g$ distributions {increase with increasing} $\xi$, though the distributions shift towards larger $x$.

\begin{figure}
	\centering
	\includegraphics[scale=0.4]{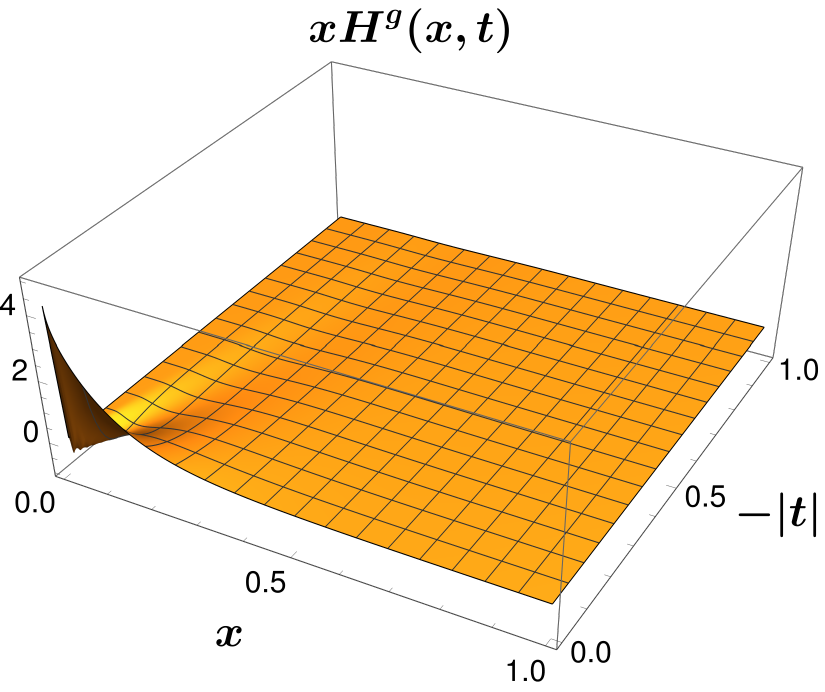}\hspace{0.2cm}
	\includegraphics[scale=0.4]{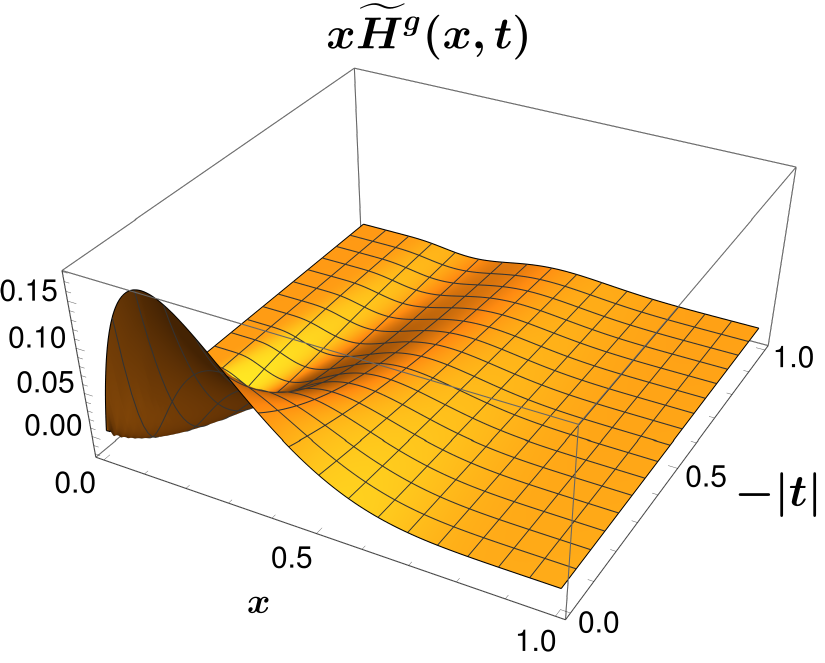}\vspace{0.2cm}
	\includegraphics[scale=0.4]{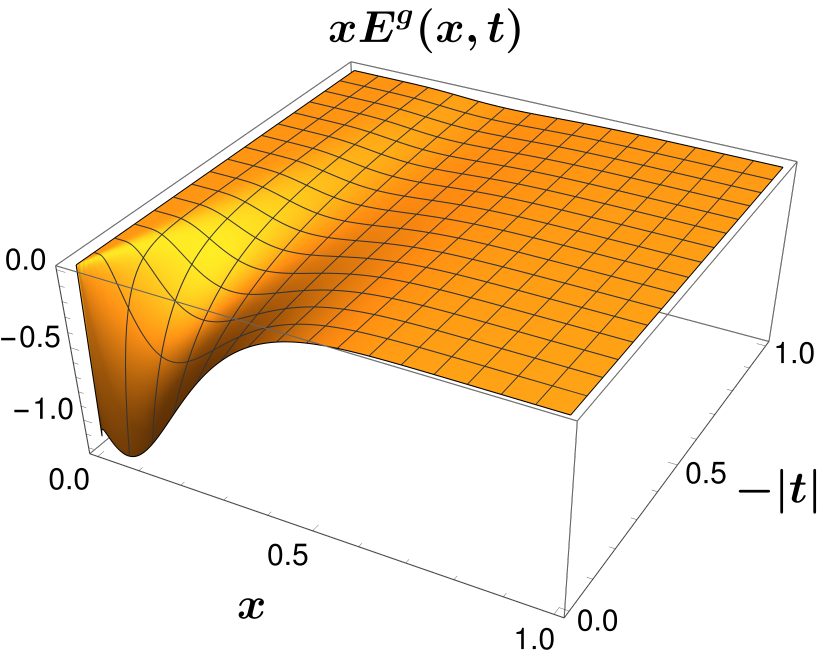}
    \vspace{0.2cm}
	\includegraphics[scale=0.4]{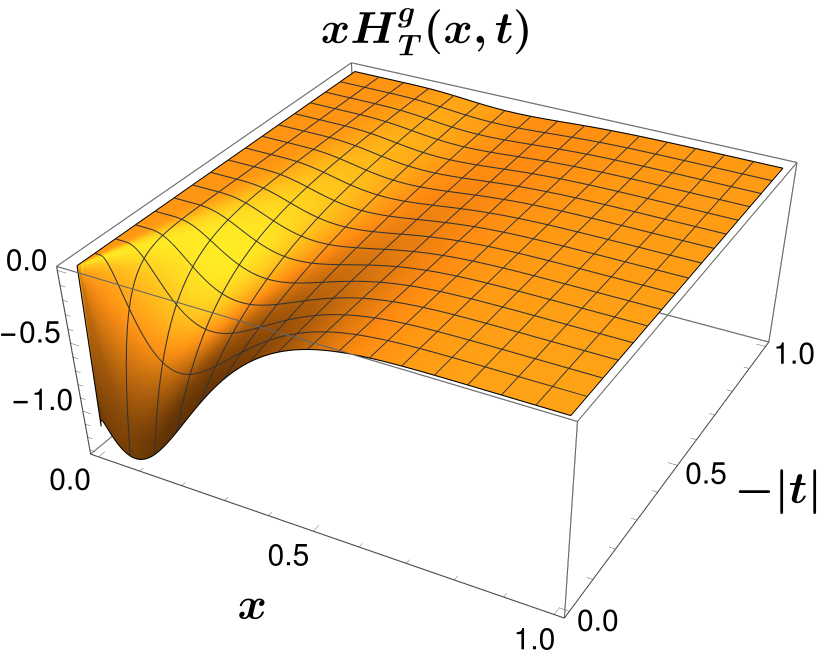}\hspace{0.5cm}
    \includegraphics[scale=0.4]{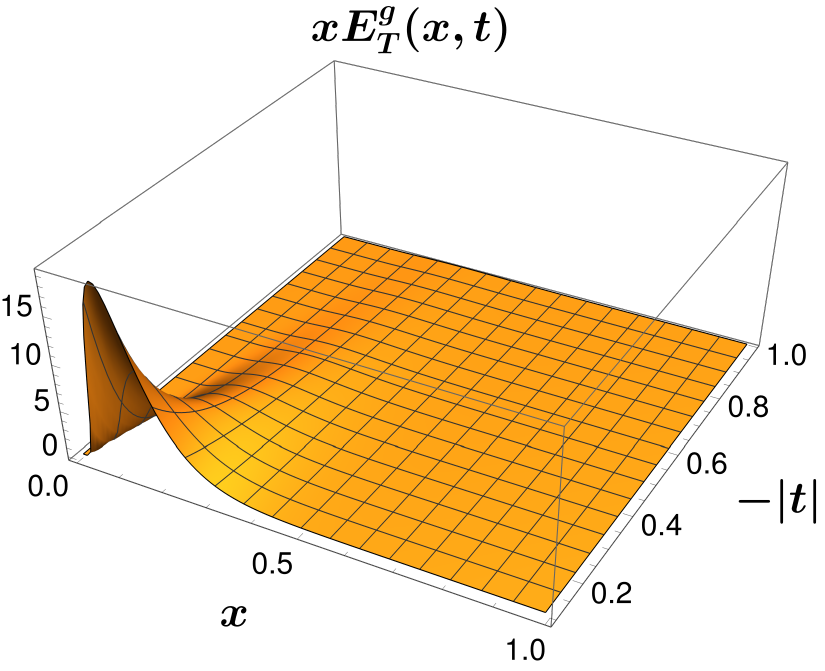}
    \caption{3D representation of gluon GPDs as a function of gluon longitudinal momentum fraction $x$ and transverse momentum transfer squared $|t|$ (in GeV$^2$) at 
    $\xi=0$.}
    \label{fig:3Dmomentum GPDS}
\end{figure}
\begin{figure}
	\centering
	\includegraphics[scale=0.4]{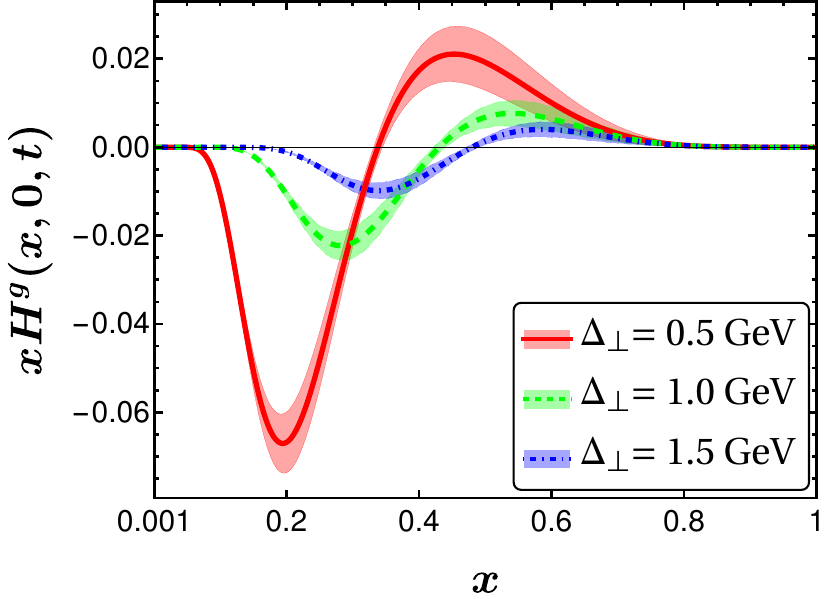}\hspace{0.2cm}
	\includegraphics[scale=0.4]{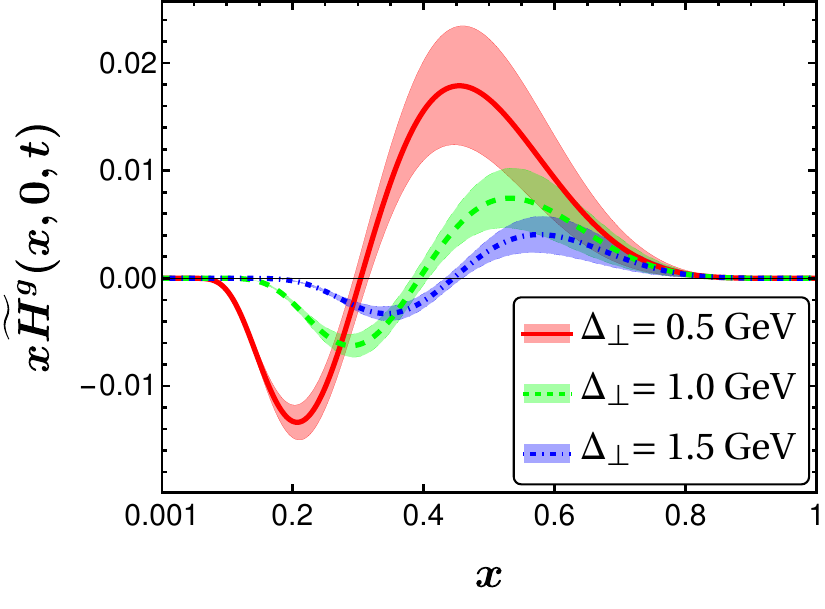}\hspace{0.2cm}
    \includegraphics[scale=0.4]{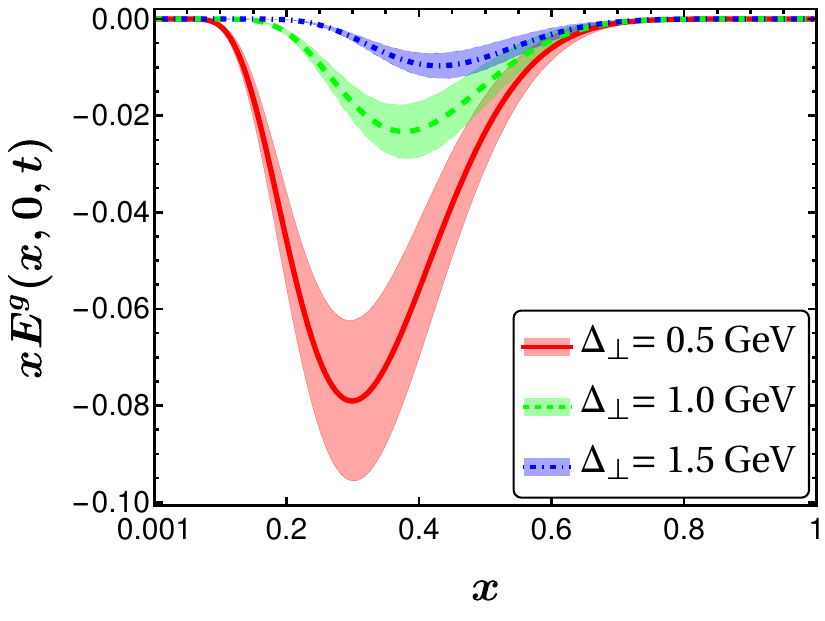}
    \vspace{0.2cm}
	\includegraphics[scale=0.4]{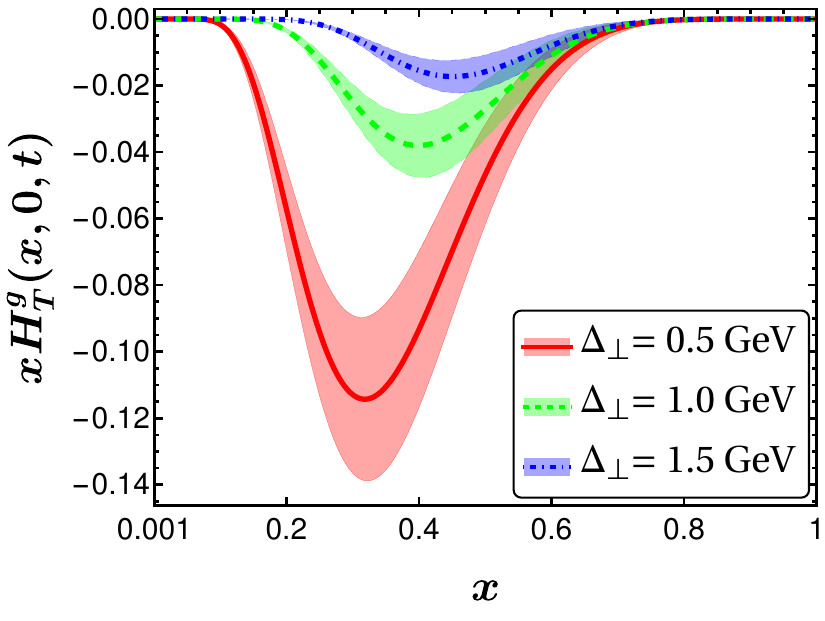}\hspace{0.2cm}
    \includegraphics[scale=0.4]{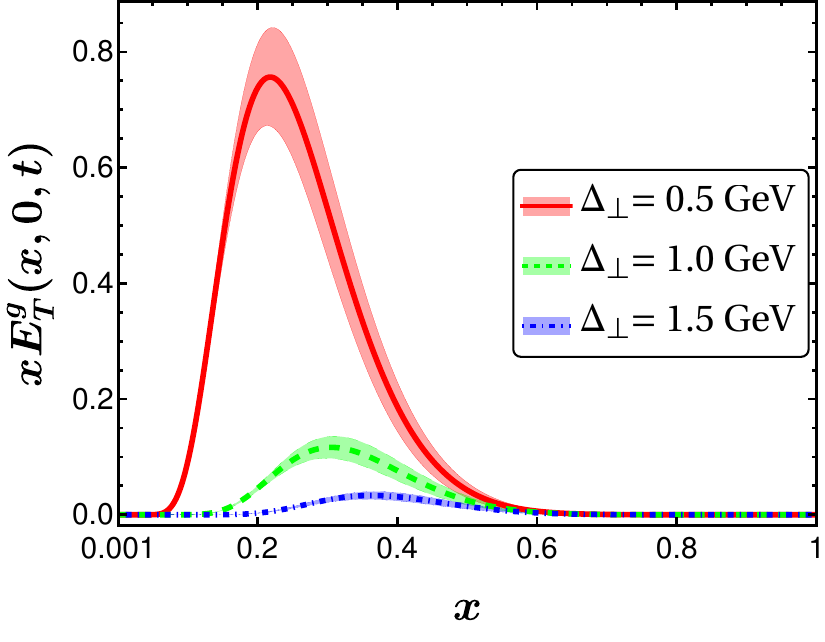}
    \caption{2D plots of gluon GPDs at $\xi=0$ as a function of longitudinal momentum fraction $x$ at certain values of transverse  momentum transfer, $\Delta_{\perp}$ (in GeV).}
   \label{fig:2Dmomentum GPDS}
\end{figure}
\begin{figure}
    \centering
    \includegraphics[scale=0.52]{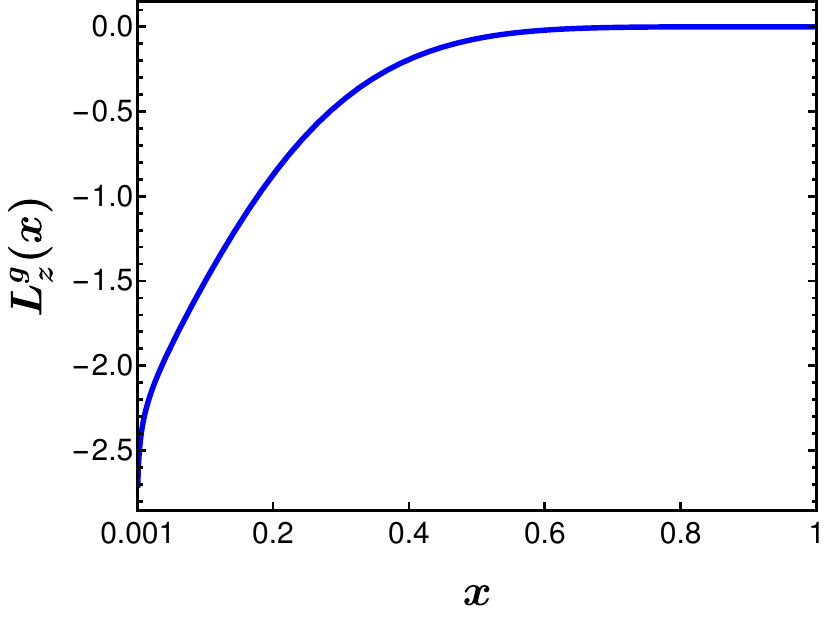}
    \caption{Model predictions on unintegrated gluon kinetic 
    OAM as a function of momentum fraction $x$ .}
    \label{fig:kineticOAM}
\end{figure}
\subsection{GPDs in $\xi=0$ limit} \label{sec:skewness zero}
In the limit $\xi \to 0 $, the correlation functions of the GPDs reduce to a simplified form, resulting in an expression solely in terms of $x$ and $t= -\Delta_\perp^2$. In the context of our specific model, we reformulate the equations depicting gluon GPDs at $\xi=0$ limit as
{\small
\begin{equation}
    \begin{aligned}
        H^g(x,0,-\Delta_\perp^2)= 2 N_g^2 x^{2b+1} (1-x)^{2 a} \Big[ \frac{1+(1-x)^2}{ x^{2}(1-x)^2} \left( \frac{\kappa^2 x^2}{\ln \frac{1}{1-x}}-(1-x)^2 \frac{\Delta_\perp^2}{4} \right)+ \left(M-\frac{M_X}{1-x} \right)^2\Big] exp\left[-\frac{\ln \left( \frac{1}{1-x}\right)(1-x)^2}{4  \kappa^2 x^2}\Delta_\perp^2\right]
        \label{eq:unpolatxizero}
    \end{aligned}
    \end{equation}}
{\small \begin{equation}
    \begin{aligned}
       \widetilde{H}^g(x,0,-\Delta_\perp^2)= 2 N_g^2 x^{2b+1} (1-x)^{2 a} \Big[ \frac{1-(1-x)^2}{x^{2} (1-x)^2} \left( \frac{\kappa^2 x^2}{\ln \frac{1}{1-x}}-(1-x)^2 \frac{\Delta_\perp^2}{4} \right)+ \left(M-\frac{M_X}{1-x} \right)^2\Big]  exp\left[-\frac{\ln \left( \frac{1}{1-x}\right)(1-x)^2}{4  \kappa^2 x^2}\Delta_\perp^2\right]
    \end{aligned}
    \label{eq:helicityatxizero}
\end{equation}}

\begin{equation}
    \begin{aligned}
       E^g(x,0,-\Delta_\perp^2)=  4 M N_g^2 x^{2b} (1-x)^{2 a} \left(M(1-x)-M_X \right) exp\left[-\frac{\ln \left( \frac{1}{1-x}\right)(1-x)^2}{4  \kappa^2 x^2}\Delta_\perp^2\right]
    \end{aligned}
    \label{eq:Egatxizero}
\end{equation}

\begin{equation}
    \begin{aligned}
       H^g_{T} (x,0,-\Delta_\perp^2)=  4 M N_g^2 x^{2b} (1-x)^{2 a-1} \left(M(1-x)-M_X \right) exp\left[-\frac{\ln \left( \frac{1}{1-x}\right) \Delta_\perp^2 (1-x)^2}{4 \kappa^2 x^2}\right]
    \end{aligned}
\end{equation}

\begin{equation}
    \begin{aligned}
       E^g_T(x,0,-\Delta_\perp^2)=  4 M^2 N_g^2 x^{2b-1} (1-x)^{2 a+1}   exp\left[-\frac{\ln \left( \frac{1}{1-x}\right)(1-x)^2}{4  \kappa^2 x^2}\Delta_\perp^2\right]
    \end{aligned}
\end{equation}

 The unpolarized gluon GPD, $H^{g}(x,\xi=0,t)$ {given in Eq.~(\ref{eq:unpolatxizero})} reduces to the unpolarized gluon parton distribution function in the forward limit, i.e., $f_{1}^{g}(x)=H^{g}(x,\xi=0,t=0)$. Similarly, the helicity-dependent gluon GPD, $\widetilde{H}^{g}(x,\xi=0,t)$  Eq.~(\ref{eq:helicityatxizero}) gives the gluon helicity pdf in the forward limit as, $g_{1L}^{g}(x)=\widetilde{H}^{g}(x,\xi=0,t=0)$. The gluon helicity GPD provides the gluon spin contribution to the total proton spin $\Delta G=\int {\rm d}x g_{1L}(x)$.

Fig.~\ref{fig:3Dmomentum GPDS} shows the 3-dimensional behavior of non-zero gluon GPDs of proton in the momentum space at zero skewness with respect to $x$ and $t=-\Delta_{\perp}^2$. 
 The unpolarized GPD $H^{g}(x,\xi=0,t)$ at $\xi=0$  shows a sharp peak near $x=0$, only for small $\Delta_{\perp}$. 
 The behaviour around $x \to 0$  drastically changes when $\Delta_{\perp} \geq 0.05$ GeV$^{2}$. It can be better understood from the first plot in Fig.~\ref{fig:2Dmomentum GPDS}. The helicity GPD, $\widetilde{H}^{g}(x,\xi=0,t)$ peaks around $x\approx0.2$ for smaller values of momentum transfer and its peak decreases a bit slowly as compared to the unpolarized GPD with increasing $x$ and  $t$. The helicity GPD $\widetilde{H}^{g}$ also shows a similar trend as the unpolarized GPD, $H^{g}$. A similar kind of behavior of gluon GPDs has been reported in the light-cone spectator model~\cite{Tan:2023kbl}. Whereas the unpolarized and helicity GPDs $H^{g}$ and $\widetilde{H}_{g}$ are found to be positive for all values of $x$ and $t$ in basis light-front quantization (BLFQ)~\cite{Lin:2023ezw} and extended holographic light-front QCD (HLFQCD)~\cite{Gurjar:2022jkx} approaches. The gluon GPD $E^g$, which is also related to the spin-flip gravitational form factor $B(Q^2)$, is negative in our model. It peaks around $t=0$, which is the forward limit of GPDs, but it does not correspond to any PDF. There are two more non-zero chiral odd gluon GPDs, $H^{g}_T$ and $E^{g}_T$. We notice that the behaviour of GPD $H^{g}_T(x,\xi=0,t)$ is  quite similar to the  GPD $E^{g}(x,\xi=0,t)$ and they are related to each other as $E^{g}=(1-x)H_{T}^{g}$. Similarly, in Ref.~\cite{Tan:2023kbl,Meissner:2007rx} the authors have derived a relation between these GPDs as $E^{g}=xH_{T}^{g}$. Both $E^{g}$ and $H_{T}^{g}$ GPDs are negative and peak around the same values of $x$ and $t$. The GPD $E^{g}_T(x,\xi=0,t)$ has the largest amplitude in the smaller value of $x$ but it also vanishes with increasing value of $x$. All the GPDs vanish as $x\to 1$, independent of the choice of the momentum transfer.

 In Fig.~\ref{fig:2Dmomentum GPDS} we show the gluon GPDs $H^{g}(x,0,t)$, $\widetilde{H}^{g}$, $E^{g}(x,0,t)$, $H^{g}_{T}(x,0,t)$ and $E^{g}_{T}(x,0,t)$ with $x$ at certain values of momentum transfer $\Delta_{\perp}=0.5,1,1.5$ GeV, respectively.

\section{Orbital angular momentum (OAM)} \label{OAM}
As discussed in the introduction, gluons contribute significantly to the spin of the nucleon. However, the decomposition of the nucleon spin into quark and gluon intrinsic spin and OAM parts is not unique. There is also a question of separation of gluon contribution into those in a gauge invariant manner. Polarised scattering experiments have measured spin asymmetries which are directly sensitive to gluon intrinsic spin. The experimental observables must be related to a gauge-invariant object. This led to a lot of theoretical discussions, a consolidated summary can be found in \cite{Leader:2013jra}.    
There are two main decompositions: kinetic and canonical. Below, we investigate both the kinetic and canonical gluon orbital angular momentum in this model.   


\subsection{Kinetic OAM }
According to Ji's sum rule~\cite{Ji:1996ek}, the total angular momentum $J^{g}_{z}$ { of the gluons} can be obtained via the moments of the chiral even helicity conserving GPD $H^{g}$ and helicity non conserving  GPD $E^{g}$ { through the following sum rule }:
\begin{eqnarray}\label{eq:totalOAM}
    J^{g}_{z}=\frac{1}{2}\int dx x\left[H^{g}(x,0,0)+E^{g}(x,0,0)\right]
\end{eqnarray}

Substituting Equations (\ref{eq:unpolatxizero}) and (\ref{eq:Egatxizero}) into Equation (\ref{eq:totalOAM}), we obtained the gluon's total angular momentum as $J_{z}^{g} = 0.058$ in our model, which aligns with recent findings from the BLFQ collaboration, where $J^{g}_{z}|_{\text{BLFQ}} = 0.066$~\cite{Lin:2023ezw}. However, it should be noted that the scale in \cite{Lin:2023ezw} is $0.5$ GeV, which is different from that used in our model.  

{ {In an analogous gluon spectator model, with a different wave function, the gluon's total angular momentum is determined as $J_{z}^{g} = 0.19$~\cite{Tan:2023kbl}. The results from this model \cite{Tan:2023kbl} are consistent with the recent lattice result $J_{z}^{g} = 0.187(46)(10)$ obtained by the ETM Collaboration~\cite{Alexandrou:2020sml}, where the lattice result is provided in the MS scheme at a scale of 2 GeV.} }

Using the sum rules of gluon  GPDs $H^{g},~\widetilde{H}^{g}~\text{and}~E^{g}$ we calculate the gluon 
OAM in the light-cone gauge from the expression~\cite{Ji:1996ek,Chen:2008ag,Wakamatsu:2010qj,Leader:2013jra}
\begin{eqnarray}\label{eq:kineticOAM}
 L_{z}^{g}=\int dx \bigg{\{}\frac{1}{2} x\big[H^{g}(x,0,0)+E^{g}(x,0,0)\big]-\widetilde{H}^{g}(x,0,0)\bigg{\}}
\end{eqnarray}
Our numerical results show that { $L^{g}_{z}=-0.42$ }which means that the gluon kinetic 
OAM is negative. In Fig.~\ref{fig:kineticOAM} we show the variation of the unintegrated gluonic kinetic 
OAM with respect to gluon longitudinal momentum fraction $x$. Furthermore, it is noteworthy to highlight that the contribution in the small-$x$ region is not negligible. The kinetic OAM distribution peaks at low $x$, decreasing rapidly as $x$ goes to 1. Similar behavior of gluon kinetic 
OAM has been reported in another gluon spectator model, with their calculated value for gluon kinetic 
OAM being $L^{g}_{z}=-0.123$~\cite{Tan:2023kbl}, which is also found to be negative.

 \subsection{Canonical OAM from GTMDs}
The gluon canonical OAM can be calculated using chirally-even gluon GTMDs in light-cone gauge \cite{Jaffe:1989jz, Wakamatsu:2010qj}. In this section, we investigate the gluon GTMDs and canonical OAM in our model. 

In order to obtain the gluon GTMDs we used the parametrization~\cite{Lorce:2013pza,Kanazawa:2014nha,Ji:2012ba,More:2017zqp} as
\begin{eqnarray}
    xW(x,\xi=0,\bfp,\Delta_{\perp})=\int \frac{dz^{-}}{2\pi}\frac{d^2z_{\perp}}{(2\pi)^{2}}e^{ip.z}\bigg\langle p+\frac{\Delta_{\perp}}{2}\bigg|\Gamma^{ij}F^{+i}\left(-\frac{z}{2}\right)F^{+j}\left(\frac{z}{2}\right)\bigg|p-\frac{\Delta_{\perp}}{2}\bigg\rangle\Bigg{|}_{z^{+}=0}
\end{eqnarray}
where we have suppressed the color indices in the GTMD correlator because we are considering the light-cone gauge in which the gauge link becomes to be unity. 
The GTMDs, $F_{1,1}$ $F_{1,4}$ describe the distortion of unpolarized partons inside a longitudinally polarized target, whereas $G_{1,1}$ describes how the longitudinally polarized parton distorts their distribution inside an unpolarized target.
 The light cone overlap representation of the chiral-even gluon GTMDs $F_{1,1}^{g}$, $F_{1,4}^{g}$, $G_{1,1}^{g}$ and $G_{1,4}^{g}$ is given by~\cite{More:2017zqp,Kanazawa:2014nha}
 \begin{align}
     F_{1,1}^{g}=&\frac{1}{2(2\pi)^{3}}\frac{1}{2}\sum_{\Lambda,\mu,\lambda}\epsilon_{\alpha}^{\mu\ast}\epsilon_{\beta}^{\mu}\psi^{\Lambda\ast}_{\mu,\lambda}(x,\bfp^{\prime}) \psi^{\Lambda}_{\mu,\lambda}(x,\bfp^{\prime \prime}) \nonumber\\
     =& \frac{1}{2(2\pi)^{3}}\sum_{\mu,\lambda}(\epsilon_{1}^{\mu\ast}\epsilon_{1}^{\mu}+\epsilon_{2}^{{\mu}\ast}\epsilon_{2}^{\mu})\bigg[\psi^{\uparrow\ast}_{\mu,\lambda}(x,\bfp^{\prime}) \psi^{\uparrow}_{\mu,\lambda}(x,\bfp^{\prime \prime})\bigg]
 \end{align}
 \begin{align}
     G_{1,4}^{g}=& -\frac{1}{2(2\pi)^{3}}i\epsilon_{T}^{\alpha\beta}\frac{1}{2}\sum_{\Lambda,\mu,\lambda}\epsilon_{\alpha}^{\mu\ast}\epsilon_{\beta}^{\mu}\psi^{\Lambda\ast}_{\mu,\lambda}(x,\bfp^{\prime}) \psi^{\Lambda}_{\mu,\lambda}(x,\bfp^{\prime \prime}) \\ \nonumber 
        =&\frac{1}{2(2\pi)^{3}}i\sum_{\mu,\lambda}(\epsilon_{2}^{{\mu}\ast}\epsilon_{1}^{\mu}-\epsilon_{1}^{{\mu}\ast}\epsilon_{2}^{\mu})\bigg[\psi^{\uparrow\ast}_{\mu,\lambda}(x,\bfp^{\prime}) \psi^{\uparrow}_{\mu,\lambda}(x,\bfp^{\prime \prime})\bigg]
 \end{align}
\begin{align}
\tfrac{i(\mathbf{p}_\perp\times\uvec\Delta_\perp)_z}{M^2}\,F_{1,4}^g&=\frac{1}{2(2\pi)^3}\,\frac{1}{2}\sum_{\Lambda,\mu,\lambda}\text{sign}(\Lambda)\left[ 
\psi^{\Lambda\ast}_{\mu,\lambda}(x,\bfp^{\prime}) \psi^{\Lambda}_{\mu,\lambda}(x,\bfp^{\prime \prime})
\right]\nonumber\\
&=\frac{i}{2(2\pi)^3}\sum_{\mu} \textrm{Im}\!\left[\psi^{\uparrow\ast}_{\mu,+\frac{1}{2}}(x,\bfp^{\prime }) \psi^{\uparrow}_{\mu,+\frac{1}{2}}(x,\bfp^{\prime \prime})\right],\\
-\tfrac{i(\mathbf{p}_\perp\times\uvec\Delta_\perp)_z}{M^2}\,G_{1,1}^g&=\frac{1}{2(2\pi)^3}\,\frac{1}{2}\sum_{\Lambda,\lambda,\mu}{\rm sign}(\mu)\left[ 
\psi^{\Lambda\ast}_{\mu,\lambda}(x,\bfp^{\prime}) \psi^{\Lambda}_{\mu,\lambda}(x,\bfp^{\prime \prime})
\right]\nonumber\\
&=\frac{i}{2(2\pi)^3}\,\sum_\mu{\rm sign}(\mu)\,\textrm{Im}\!\left[\psi^{\uparrow\ast}_{\mu, +\frac{1}{2}}(x,\bfp^{\prime }) \psi^{\uparrow}_{\mu, +\frac{1}{2}}(x,\bfp^{\prime \prime}) \right] 
\end{align}
where $\Lambda$, $\lambda$ and $\mu$ denoted the proton, quark and gluon helicities, respectively. Also $ \bfp^{\prime \prime }$ and $ \bfp^{\prime}$ are gluon transverse momentum in the proton initial and final state, given as follow:
\be
\bfp^{\prime}=\bfp+(1-x)\frac{\bf{\Delta}_{\perp}}{2}, \hspace{0.5cm} \bfp^{\prime \prime}=\bfp-(1-x)\frac{\bf{\Delta}_{\perp}}{2}.
\ee

By employing the proton LFWFs from Eqs.~(\ref{LFWFsuparrow}) and (\ref{LFWFsdownarrow}), the analytical expressions for the above four chiral-even GTMDs can be obtained as, 
\begin{align}\label{eq:F11_GTMD}\nonumber
    F_{1,1}(x,\xi=0,\bfp,\Delta_{\perp})=\frac{2N_{g}^{2}}{{\pi  \kappa ^2}} x^{2 b-1}&(1-x)^{2 a} \log \left(\frac{1}{1-x}\right) \Bigg\{\frac{\left(1+(1-x)^2\right) \left(\bfp^2-(1-x)^2\frac{\Delta_{\perp}^2}{4} \right)}{x^2 (1-x)^2}+\left(M-\frac{M_{X}}{1-x}\right)^2 \\
    &+\frac{i (1-(1-x)^{2}) (1-x)\left(\bfp\times\Delta_{\perp}\right)}{x^{2} (1-x)^{2}}\Bigg\} \exp \Bigg[-\frac{\log \left(\frac{1}{1-x}\right) }{\kappa ^2 x^2}\left(\bfp^2+(1-x)^2\frac{\Delta_{\perp} ^2}{4}  \right)\Bigg] 
\end{align}
\begin{align}\label{eq:G14_GTMD}\nonumber
    G_{1,4}(x,\xi=0,\bfp,\Delta_{\perp})=\frac{2N_{g}^{2}}{{\pi  \kappa ^2}} x^{2 b-1}&(1-x)^{2 a} \log \left(\frac{1}{1-x}\right) \Bigg\{\frac{\left(1-(1-x)^2\right) \left(\bfp^2-(1-x)^2\frac{\Delta_{\perp}^2}{4}\right)}{x^2 (1-x)^2}+\left(M-\frac{M_{X}}{1-x}\right)^2 \\
    &+\frac{i (1+(1-x)^2) (1-x)\left(\bfp\times\Delta_{\perp}\right)}{x^{2} (1-x)^{2}}\Bigg\} \exp \Bigg[-\frac{\log \left(\frac{1}{1-x}\right) }{\kappa ^2 x^2}\left(\bfp^2+(1-x)^2\frac{\Delta_{\perp} ^2}{4}  \right)\Bigg] 
\end{align}
\begin{align}\label{eq:F14_GTMD}
	F_{1,4}(x,\xi=0,\bfp,\Delta_{\perp})=\frac{{ 2} N_{g}^{2}M^{2}}{{\pi  \kappa ^2}}x^{2 b-1}(1-x)^{2 a+1} & \log \left(\frac{1}{1-x}\right) \left\{\frac{\left(1-(1-x)^2\right)}{x^2 (1-x)^2}\right\} \nonumber \\
	& \times \exp \left[-\frac{\log \left(\frac{1}{1-x}\right) }{\kappa ^2 x^2}\left(\bfp^2+(1-x)^2\frac{\Delta_{\perp} ^2}{4}  \right)\right] 
\end{align}

\begin{align}\label{eq:G11_GTMD}
	G_{1,1}(x,\xi=0,\bfp,\Delta_{\perp})=-\frac{{ 2} N_{g}^{2}M^{2}}{{\pi  \kappa ^2}}x^{2 b-1}(1-x)^{2 a+1} & \log \left(\frac{1}{1-x}\right) \left\{\frac{\left(1+(1-x)^2\right)}{x^2 (1-x)^2}\right\} \nonumber \\
	& \times \exp \left[-\frac{\log \left(\frac{1}{1-x}\right) }{\kappa ^2 x^2}\left(\bfp^2+(1-x)^2\frac{\Delta_{\perp} ^2}{4}  \right)\right] 
\end{align}

In the forward limit, i.e., at $\xi=0$ and $\Delta_{\perp}=0$ limit the unpolarized gluon GTMD, $F_{1,1}^{g}$ gives the unpolarized gluon TMD which describe the unpolarized gluon density as,
\begin{eqnarray}
    f_{1}^{g}(x)=\int d^{2}\bfp F_{1,1}^{g}(x,0,\bfp,0,0)
\end{eqnarray}
which we discussed in our previous work~\cite{Chakrabarti:2023djs}. Whereas, the canonical orbital angular momentum of gluon, $\ell_{z}^{g}$ can be obtained from GTMD $F_{1,4}^{g}$ in the light-cone gauge as~\cite{Bhattacharya:2022vvo,Lorce:2011kd,Hatta:2011ku,Leader:2013jra,Singh:2023hgu},
\begin{eqnarray}
    \ell_{z}^{g}(x)=-\int d^{2}\bfp\frac{\bfp^{2}}{M^{2}}F_{1,4}^{g}(x,0,\bfp,0,0).
\end{eqnarray}
The $x$ dependence of gluon canonical orbital angular momentum, $\ell_{z}^{g}$ can be given as,
\begin{eqnarray}
    \ell^{g}_z(x)=-2N_{g}^2 \kappa^2 \frac{1-(1-x)^2}{x^2 (1-x)^2} x^{2 b +3}(1-x)^{2 a+1} \frac{1}{\log [\frac{1}{1-x}]}
    \end{eqnarray}
    \begin{figure}
    \centering
    \includegraphics[scale=0.52]{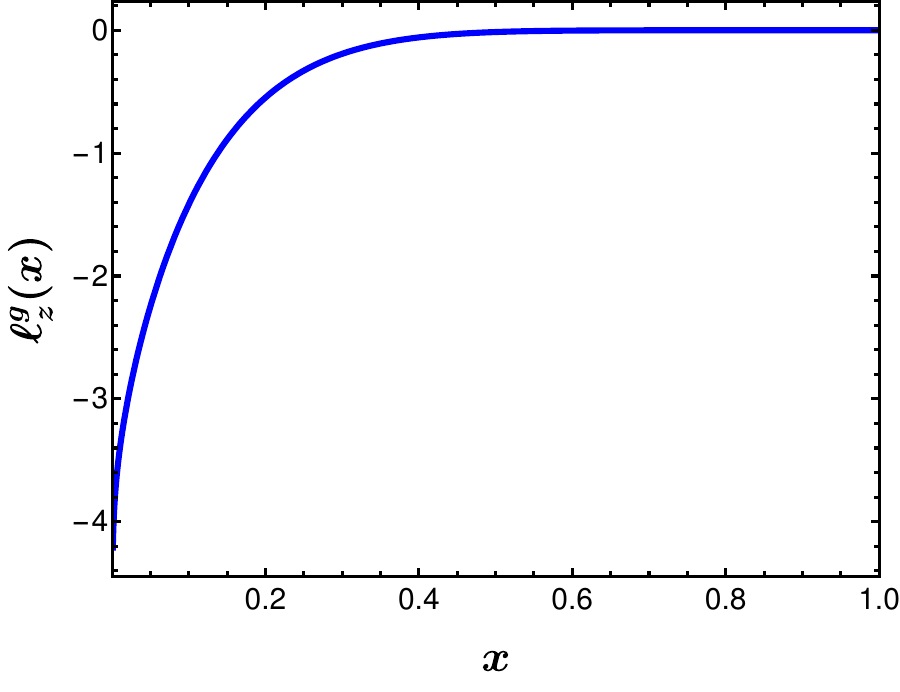}\hspace{0.5cm}
    \includegraphics[scale=0.52]{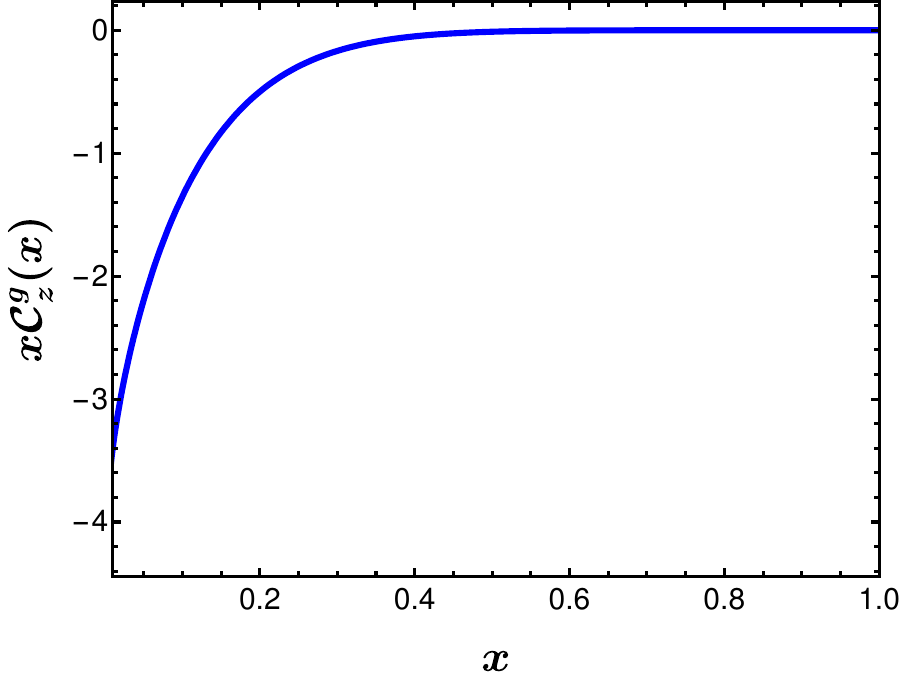}
    \caption{Model predictions for the gluon canonical OAM which can be obtained with $F_{1,4}^{g}$ gluon GTMD (Left) whereas the gluon spin-orbit correlations function which can be obtained with gluon $G_{1,1}^{g}$ GTMD (Right). }
    \label{fig:OAMs}
\end{figure}
Similarly, the spin-orbit correlation factor for the gluons can be obtained by using the gluon GTMD $G_{1,1}^{g}$ as~\cite{Chakrabarti:2017teq,Lorce:2011kd,Maji:2022tog,Gurjar:2021dyv},
\begin{eqnarray}
    C_{z}^{g}(x)=\int d^{2}\bfp \frac{\bfp^{2}}{M^{2}}G_{1,1}^{g}(x,0,\bfp,0,0)
\end{eqnarray}
and the fourth chiral even GTMD $G_{1,4}^{g}$ gives the gluon helicity TMD $g_{1L}^{g}$ in the forward limit, corresponding collinear parton distribution is gluon helicity PDF which contributes to the proton spin as,
\begin{eqnarray}
    \Delta G=\int dx d^{2}\bfp G_{1,4}^{g}(x,0,\bfp,0,0)
\end{eqnarray}
In Fig.~\ref{fig:OAMs} we show the gluon momentum fraction $x$ dependence of canonical orbital angular momentum in the left panel, whereas the spin-orbit correlation function has been depicted in the right panel. One can notice that the gluon canonical OAM, $\ell_{z}^{g}(x)$ and spin-orbit correlation factor $\mathcal{C}_{z}^{g}(x)$ both {distributions have negative values} in the whole range of $x$.  After integrating $\ell_{z}^{g}(x)$ over $x$ one can obtain the numerical value of gluon canonical orbital angular momentum which gives the contribution into Jaffe–Manohar spin sum rule. In our model calculations, we obtained the numerical value of canonical OAM as { $\ell_{z}^{g}=-0.38$}, { which is in good agreement with the another spectator model canonical OAM result, $\ell^{g}_{z}\simeq -0.333$~\cite{Tan:2023vvi}}. As similar to quarks, the canonical and kinetical gluon OAMs are 
comparable in our model, i.e., $\ell_{z}^{g}\sim L_{z}^{g}$ as reported in Refs.~\cite{Lorce:2012ce,Ji:2012sj,Leader:2013jra}. The spin-orbit correlation factor is { $c_{z}^{g}=-15.5$ } in our model, { the negative sign implies that the gluon spin and OAM are oriented in opposite directions}.    
\section{GPDs in impact parameter space (IPDs)} \label{sec:IPDs}
The two-dimensional Fourier transform of the GPDs at zero skewness with respect to the transverse momentum transferred, $\Delta_{\perp}$ to the process is used to determine the impact parameter-dependent generalized parton distributions, also known as IPDs~\cite{Burkardt:2002hr},
\begin{figure}
	\centering
	\includegraphics[scale=0.4]{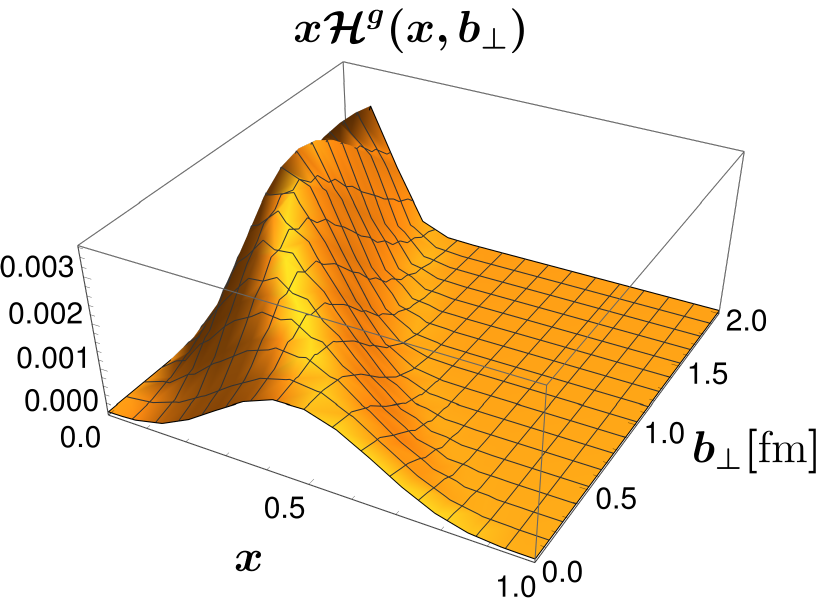}\hspace{0.2cm}
	\includegraphics[scale=0.4]{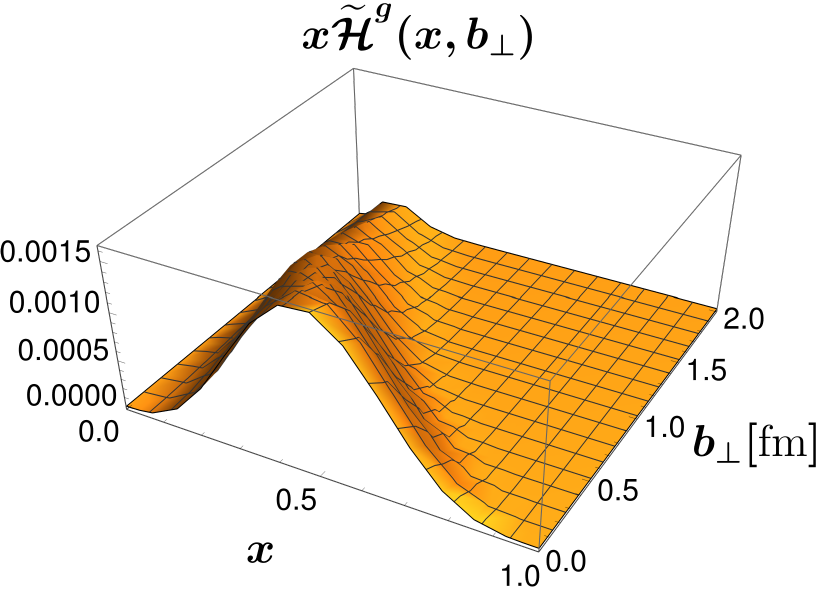} \hspace{0.2cm}
    \includegraphics[scale=0.4]{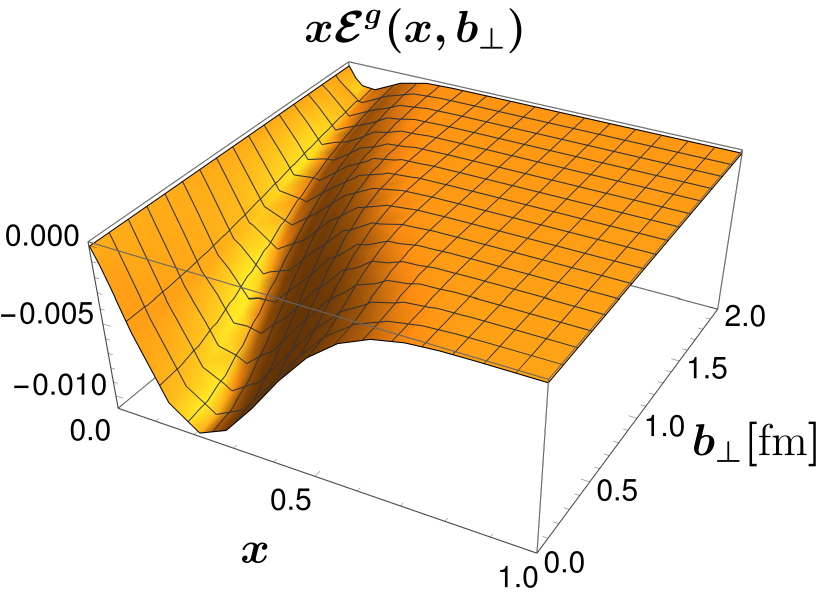} 
    \vspace{0.2cm}
    \includegraphics[scale=0.4]{ 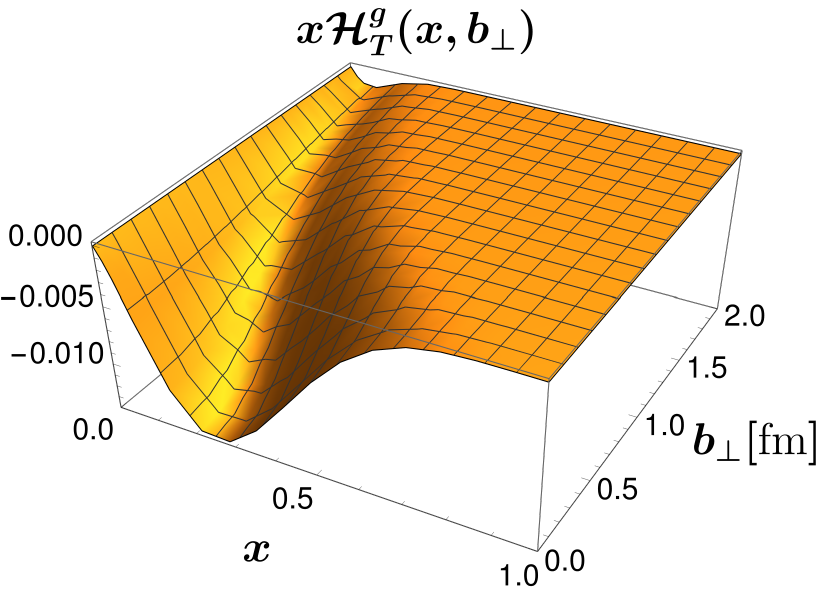}\vspace{0.2cm}
	\includegraphics[scale=0.4]{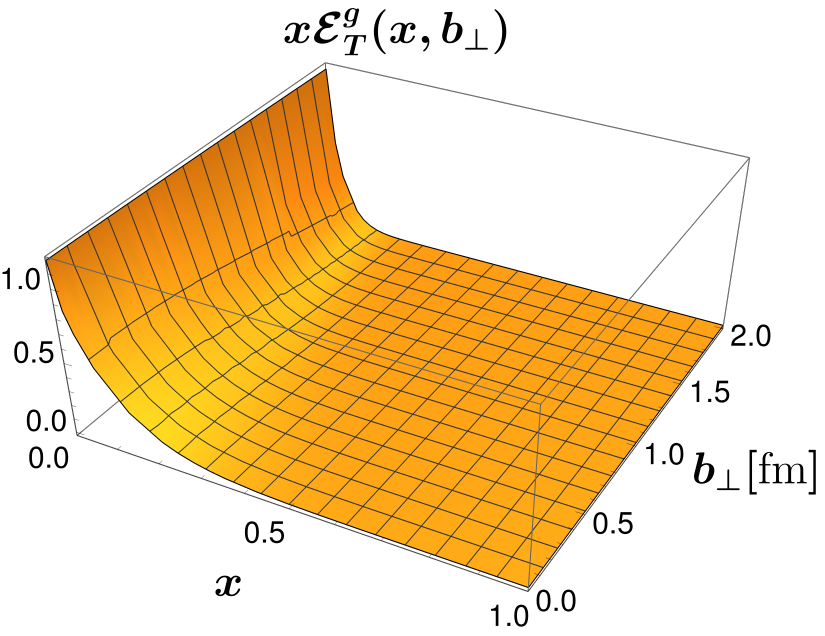}
 \caption{
 3D plots of impact parameter dependent GPDs, $\mathcal{H}^{g}$, $\widetilde{\mathcal{H}}^{g}$, $\mathcal{E}^{g}$, $\mathcal{H}_{T}^{g}$ and $\mathcal{E}_{T}^{g}$ as a function of $x$ and the transverse impact parameter $b_{\perp}$ (in fm) at $\xi=0$.} 
 \label{fig:IPDs3D}
\end{figure}

\begin{eqnarray}
\mathcal{F}(x,b_\perp)=\int \frac{d^2 \Delta_\perp}{(2 \pi)^2} e^{-i \Delta_\perp. b_\perp} F^g(x,\xi=0,t=-\Delta_\perp^2)
\end{eqnarray}
 where $b_\perp$ is the impact parameter in the transverse plane which is 
 the transverse distance between the struck parton and the center of momentum of the hadron. In the context of momentum space, the generalized parton distributions are depicted as off-diagonal matrix elements devoid of any immediate probabilistic implications. Conversely, when analyzed within the framework of impact parameter space, these distributions not only adhere to rigorous positivity conditions but also offer a compelling and substantial probabilistic interpretation~\cite{Burkardt:2000za,Burkardt:2002hr}.
 
\begin{figure}
	\centering
	\includegraphics[scale=0.4]{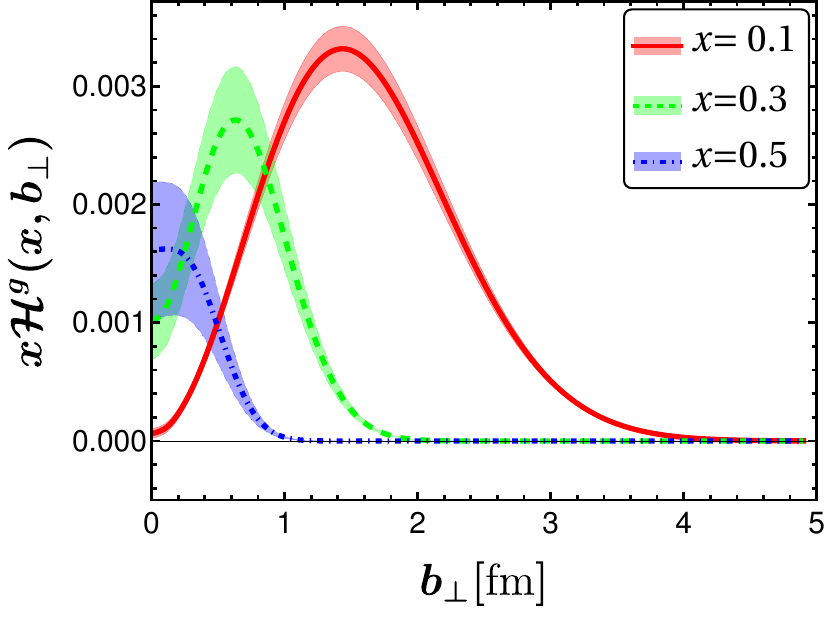}\hspace{0.2cm}
	\includegraphics[scale=0.42]{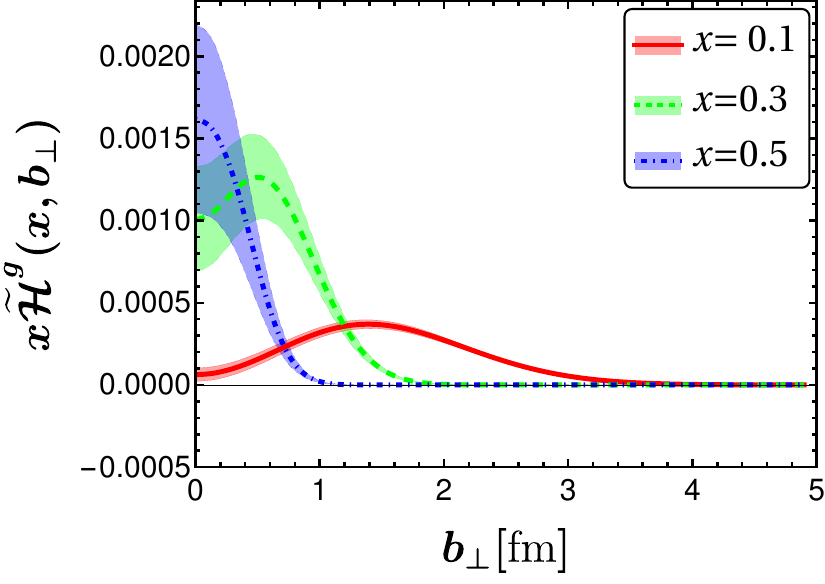}\hspace{0.2cm}
    \includegraphics[scale=0.4]{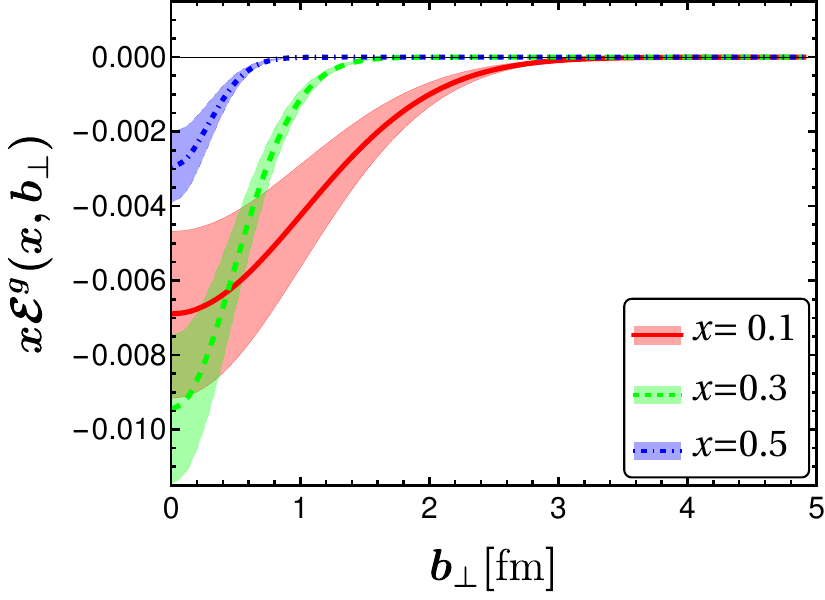} 
    \vspace{0.2cm}
    \includegraphics[scale=0.42]{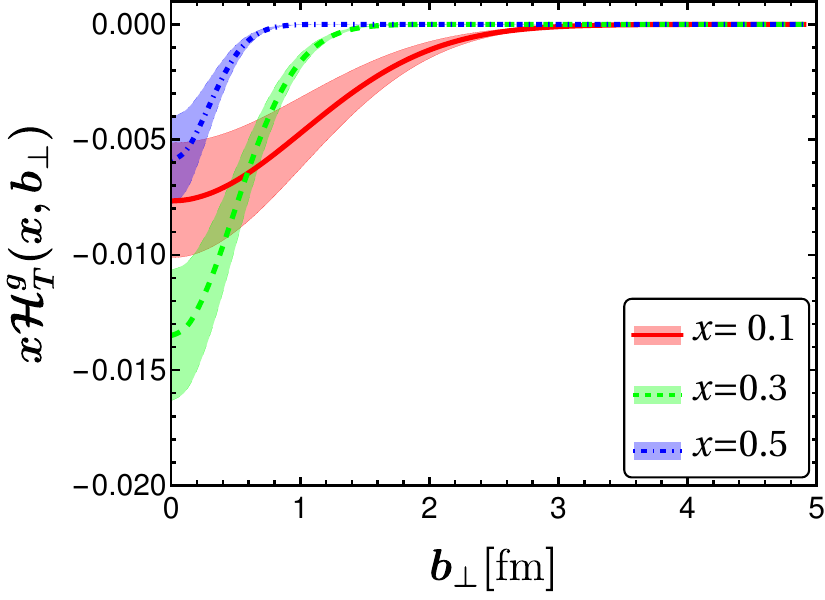}\hspace{0.2cm}
	\includegraphics[scale=0.4]{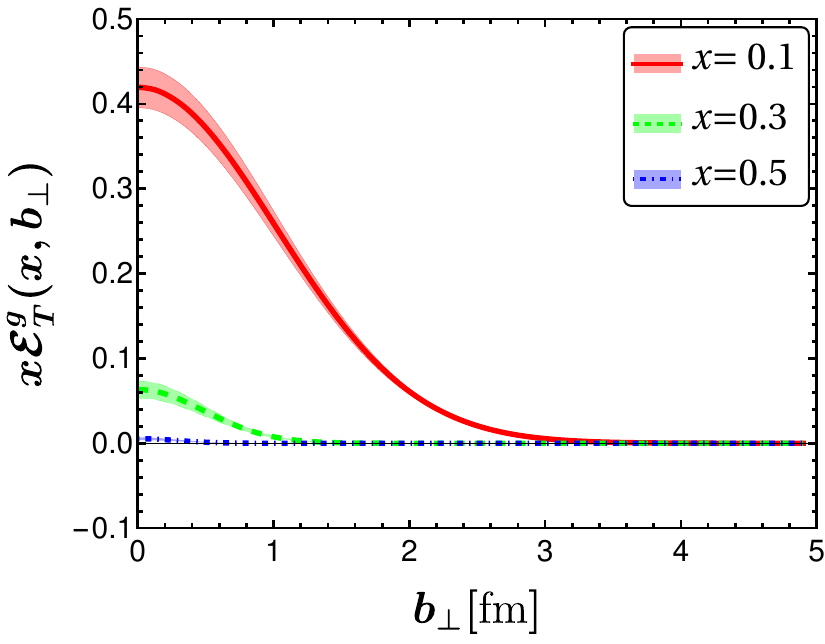}
 \caption{
 2D plots of impact parameter dependent GPDs, $\mathcal{H}^{g}$, $\widetilde{\mathcal{H}}^{g}$, $\mathcal{E}^{g}$, $\mathcal{H}_{T}^{g}$ and $\mathcal{E}_{T}^{g}$ as a function of the transverse impact parameter $b_{\perp}$ (in fm) 
 for specific values of the momentum fraction $x$, precisely, $x=0.1,~x=0.3~\text{and}~x=0.5$, with 
 $\xi=0$.}
 \label{fig:IPDs2D}
\end{figure}

 { In Figure~\ref{fig:IPDs3D}, we depict three-dimensional representations of the non-zero generalized parton distributions in impact parameter space. Both the chiral even and the chiral odd GPDs are presented as functions of gluon longitudinal momentum fraction $x$ and impact parameter $b_{\perp}$ at zero skewness.} Notably, the first plot in Figure~\ref{fig:IPDs3D} reveals that the impact parameter distribution $\mathcal{H}^{g}$ satisfies stringent positivity constraints, i.e., $\mathcal{H}^{g}\geq 0$, thereby affording a probabilistic interpretation. For a more precise understanding, one can examine the first plot in Figure~\ref{fig:IPDs2D}, which illustrates how the unpolarized IPDs change with respect to the impact parameter $b_{\perp}$ at particular values of gluon momentum fraction $x$.
Likewise, the helicity GPD in impact parameter space, $\widetilde{\mathcal{H}}^{g}$, exhibits a positive distribution across the entire transverse impact parameter range, adhering to the positivity constraint. This specific representation of the helicity GPD in impact parameter space quantifies the difference in density between gluons with positive and negative helicity. The generalized parton distributions $\mathcal{E}^{g}$ and $\mathcal{H}_{T}^{g}$ exhibit comparable shapes and display a negative distribution across the entire range of impact parameter space. To establish a probabilistic understanding, it is crucial to focus on amplitudes where the initial and final states possess matching helicities. Nevertheless, a challenge arises when attempting to develop a probabilistic interpretation for $E^{g}$ in momentum space, as it is associated with states that have differing helicities between the initial and final states. Finally, the chiral-odd GPD, denoted as $\mathcal{E}^{g}_{T}$, exhibits a positive behavior across the entire range of $b_{\perp}$. It demonstrates significant amplitude at small $x$ values but is noticeably attenuated in the region $x\geq0.5$.
\begin{figure}
    \centering
    \includegraphics[scale=0.52]{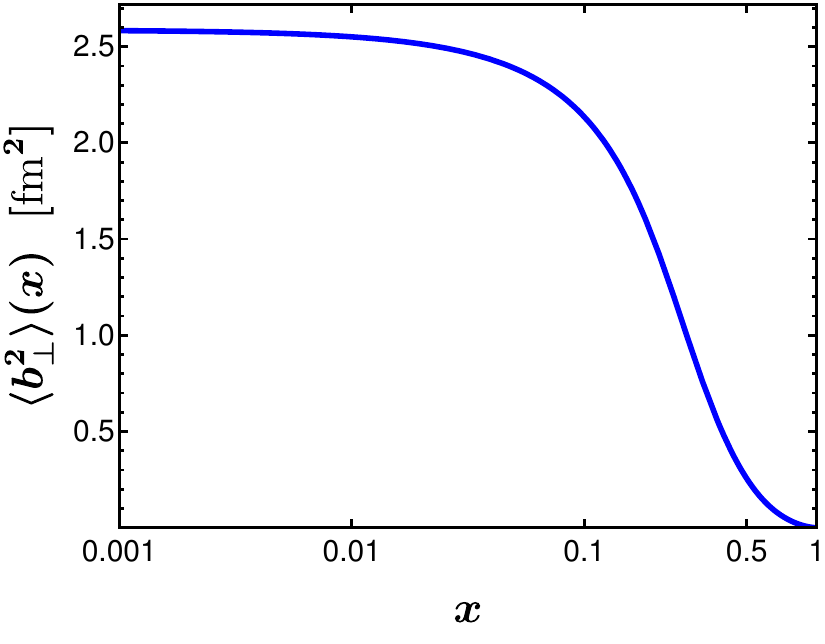}\hspace{0.5cm}
    \caption{Model predictions for the average squared transverse radius of the gluon density, denoted as $\langle b^2_{\perp}\rangle$ (in fm$^2$) as a function of $x$.}
    \label{fig:squareradii}
\end{figure}
In Fig.~\ref{fig:squareradii} we present the squared radius of gluon densities in the transverse plane { as a function of $x$}. The parameter $\langle b_{\perp}^{2}\rangle$ signifies the transverse size of the hadron, revealing an expansion of the transverse radius as parton momentum fraction $x$ decreases~\cite{Dupre:2016mai}. At lower values of $x$, the gluon displays a larger transverse average radii in contrast to the quark. Conversely, as $x$ values increase, the transverse size of the gluon decreases. As expected, when $x$ approaches to 1, the proton's transverse size resembles that of a point-like object, illustrating the color transparency of the proton~\cite{Brodsky:2022bum,Dupre:2016mai}.

\section{conclusion}\label{sec:conclusion}
In this work, we have employed a recently developed light-front spectator model, incorporating the gluon as an active parton and the rest as a spectator. The proton light-front wavefunctions are adopted from the soft-wall AdS/QCD prediction. The model provides valuable insights into the leading twist gluon TMDPDFs, GPDs, and GTMDs. Within the light-cone formalism, the above distribution functions can be expressed as the overlap of the proton wave functions. We have extended our investigation to the gluon  GPDs at zero and non-zero skewness { in the DGLAP region}, shedding light on their intricate behavior in various momentum configurations.  We have explored both chiral even and chiral-odd gluon GPDs at leading twist. At non-zero skewness, among the eight leading-twist GPDs only six of them survive in our model in which unpolarized and helicity GPD exhibit oscillatory behavior, $E^g$, $\tilde{E}^g$ and $\tilde{H}_T^g$ show negative magnitude while $E_T^g$ shows the positive distribution in the entire range of $x $ and $\xi$. We have particularly focused on the $\xi=0$ behavior of the GPDs in the momentum and impact parameter space. We have evaluated the canonical and kinetic OAMs and the spin-orbit correlations of the gluons through GTMDs and GPDs in this model dominating the small $x$ regions. We found that the canonical and kinetic orbital angular momentums are negative in the entire range of $x$ and they are approximately close to each other, as $\ell_{z}^{g}\simeq -0.38$ and $L_{z}^{g}\simeq-0.42$ in our model.
Using Ji's spin sum rule we have also calculated the total angular momentum, $J_{z}^{g}=0.058$, which agrees with the BLFQ predictions. The spin-orbit correlations are also negative, indicating that the gluon spin and OAM are anti-aligned. Further, we have also computed the average transverse square radius, $\langle b_{\perp}^{2}\rangle^{g}$ as a function of gluon longitudinal momentum fraction $x$. As expected, the transverse size decreases with increasing $x$. As $x$ approaches 1, the transverse size of the proton diminishes, and it behaves like a point-like object. Our investigation has the potential to establish important theoretical limitations regarding gluon GPDs and angular momentum. However, to confirm these predictions, further experimental measurements are essential.
\section{acknowledgement}
{ A. M. would like to thank SERB MATRICS (MTR/2021/000103) for funding.  The work of T. M. is supported by the Science and Engineering Research Board (SERB) through the SRG (Start-up Research Grant) of File No. SRG/2023/001093. The work of C. M. is supported by new faculty start up funding by the Institute of Modern Physics, Chinese Academy of Sciences, Grant No. E129952YR0.  C. M. also thanks the Chinese Academy of Sciences Presidents International Fellowship Initiative for the support via Grants No. 2021PM0023.
 }

\appendix
\section{Appendix} \label{append} 
The matrix elements $ T^g_{i}$ for the chiral even GPDs are given by the following relations with the LFWFs:

\begin{eqnarray}\label{eq:T1g}
     T^g_1 &= & \int \frac{d^2 \textbf{p}_\perp}{16 \pi^3} \Big[ \psi_{+1 +\frac{1}{2}}^{\uparrow \ast}\left(x^\prime, \bfp^{\prime} \right) \psi_{+1 +\frac{1}{2}}^{\uparrow }\left(x^{\prime \prime}, \bfp^{\prime \prime} \right) + \psi_{+1 -\frac{1}{2}}^{\uparrow \ast}\left(x^\prime, \bfp^{\prime} \right) \psi_{+1 -\frac{1}{2}}^{\uparrow }\left(x^{\prime \prime}, \bfp^{\prime \prime} \right) \nonumber \\
   & + &\psi_{+1 +\frac{1}{2}}^{\downarrow \ast}\left(x^\prime, \bfp^{\prime}\right) \psi_{+1 +\frac{1}{2}}^{\downarrow }\left(x^{\prime \prime}, \bfp^{\prime \prime} \right)+\psi_{+1 -\frac{1}{2}}^{\downarrow \ast}\left(x^\prime, \bfp^{\prime} \right) \psi_{+1 -\frac{1}{2}}^{\downarrow }\left(x^{\prime \prime}, \bfp^{\prime \prime} \right) \Big] 
\end{eqnarray}
\begin{eqnarray}
     T^g_2 &= & \int \frac{d^2 \textbf{p}_\perp}{16 \pi^3} \Big[ \psi_{+1 +\frac{1}{2}}^{\uparrow \ast}\left(x^\prime, \bfp^{\prime} \right) \psi_{+1 +\frac{1}{2}}^{\uparrow }\left(x^{\prime \prime}, \bfp^{\prime \prime} \right) + \psi_{+1 -\frac{1}{2}}^{\uparrow \ast}\left(x^\prime, \bfp^{\prime} \right) \psi_{+1 -\frac{1}{2}}^{\uparrow }\left(x^{\prime \prime}, \bfp^{\prime \prime} \right) \nonumber \\
   & - &\psi_{+1 +\frac{1}{2}}^{\downarrow \ast}\left(x^\prime, \bfp^{\prime}\right) \psi_{+1 +\frac{1}{2}}^{\downarrow }\left(x^{\prime \prime}, \bfp^{\prime \prime} \right)-\psi_{+1 -\frac{1}{2}}^{\downarrow \ast}\left(x^\prime, \bfp^{\prime} \right) \psi_{+1 -\frac{1}{2}}^{\downarrow }\left(x^{\prime \prime}, \bfp^{\prime \prime} \right) \Big] 
\end{eqnarray}
\begin{eqnarray}
T^g_3  &=& \int \frac{d^2 \textbf{p}_\perp}{16 \pi^3} \Big[ \psi_{+1 +\frac{1}{2}}^{\uparrow \ast}\left(x^\prime, \bfp^{\prime} \right) \psi_{+1 +\frac{1}{2}}^{\downarrow }\left(x^{\prime \prime}, \bfp^{\prime \prime}\right) + \psi_{+1 -\frac{1}{2}}^{\uparrow \ast}\left(x^\prime, \bfp^{\prime} \right) \psi_{+1 -\frac{1}{2}}^{\downarrow }\left(x^{\prime \prime}, \bfp^{\prime \prime} \right) \nonumber \\
     & &+\psi_{-1 +\frac{1}{2}}^{\uparrow \ast}\left(x^\prime, \bfp^{\prime}\right) \psi_{-1 +\frac{1}{2}}^{\downarrow }\left(x^{\prime \prime}, \bfp^{\prime \prime}\right)+\psi_{-1 -\frac{1}{2}}^{\uparrow \ast}\left(x^\prime, \bfp^{\prime} \right) \psi_{-1 -\frac{1}{2}}^{\downarrow }\left(x^{\prime \prime}, \bfp^{\prime \prime} \right) \Big]     
\end{eqnarray}
\begin{eqnarray}
T^g_4  &=& \int \frac{d^2 \textbf{p}_\perp}{16 \pi^3} \Big[ \psi_{+1 +\frac{1}{2}}^{\uparrow \ast}\left(x^\prime, \bfp^{\prime} \right) \psi_{+1 +\frac{1}{2}}^{\downarrow }\left(x^{\prime \prime}, \bfp^{\prime \prime}\right) + \psi_{+1 -\frac{1}{2}}^{\uparrow \ast}\left(x^\prime, \bfp^{\prime} \right) \psi_{+1 -\frac{1}{2}}^{\downarrow }\left(x^{\prime \prime}, \bfp^{\prime \prime} \right) \nonumber \\
     & &-\psi_{-1 +\frac{1}{2}}^{\uparrow \ast}\left(x^\prime, \bfp^{\prime}\right) \psi_{-1 +\frac{1}{2}}^{\downarrow }\left(x^{\prime \prime}, \bfp^{\prime \prime}\right)-\psi_{-1 -\frac{1}{2}}^{\uparrow \ast}\left(x^\prime, \bfp^{\prime} \right) \psi_{-1 -\frac{1}{2}}^{\downarrow }\left(x^{\prime \prime}, \bfp^{\prime \prime} \right) \Big]     
\end{eqnarray}
Likewise, the matrix elements $ \tilde{T}^g_{i}$ for the chiral odd GPDs in terms of LFWFs could be expressed as following: 
\begin{eqnarray}
    \widetilde{T}^g_{1}& =&  \int \frac{d^2 \textbf{p}_\perp}{16 \pi^3} \Big[ \psi_{+1 +\frac{1}{2}}^{\uparrow \ast}\left(x^\prime, \bfp^{\prime} \right) \psi_{-1 +\frac{1}{2}}^{\downarrow }\left(x^{\prime \prime}, \bfp^{\prime \prime} \right) + \psi_{+1 -\frac{1}{2}}^{\uparrow \ast}\left(x^\prime, \bfp^{\prime} \right) \psi_{-1 -\frac{1}{2}}^{\downarrow }\left(x^{\prime \prime}, \bfp^{\prime \prime} \right) \nonumber \\
  &  + &\psi_{+1 +\frac{1}{2}}^{\downarrow \ast}\left(x^\prime, \bfp^{\prime}\right) \psi_{-1 +\frac{1}{2}}^{\uparrow }\left(x^{\prime \prime}, \bfp^{\prime \prime} \right)+\psi_{+1 -\frac{1}{2}}^{\downarrow \ast}\left(x^\prime, \bfp^{\prime} \right) \psi_{-1 -\frac{1}{2}}^{\uparrow }\left(x^{\prime \prime}, \bfp^{\prime \prime} \right) \Big] 
\end{eqnarray}
\begin{eqnarray}
    \widetilde{T}^g_{2} &= & \int \frac{d^2 \textbf{p}_\perp}{16 \pi^3} \Big[ \psi_{+1 +\frac{1}{2}}^{\uparrow \ast}\left(x^\prime, \bfp^{\prime} \right) \psi_{-1 +\frac{1}{2}}^{\downarrow }\left(x^{\prime \prime}, \bfp^{\prime \prime} \right) + \psi_{+1 -\frac{1}{2}}^{\uparrow \ast}\left(x^\prime, \bfp^{\prime} \right) \psi_{-1 -\frac{1}{2}}^{\downarrow }\left(x^{\prime \prime}, \bfp^{\prime \prime} \right) \nonumber \\
   & - &\psi_{+1 +\frac{1}{2}}^{\downarrow \ast}\left(x^\prime, \bfp^{\prime}\right) \psi_{-1 +\frac{1}{2}}^{\uparrow }\left(x^{\prime \prime}, \bfp^{\prime \prime} \right)-\psi_{+1 -\frac{1}{2}}^{\downarrow \ast}\left(x^\prime, \bfp^{\prime} \right) \psi_{-1 -\frac{1}{2}}^{\uparrow }\left(x^{\prime \prime}, \bfp^{\prime \prime} \right) \Big] 
\end{eqnarray}
\begin{eqnarray}
    \widetilde{T}^g_{3} &= & \int \frac{d^2 \textbf{p}_\perp}{16 \pi^3} \Big[ \psi_{+1 +\frac{1}{2}}^{\uparrow \ast}\left(x^\prime, \bfp^{\prime} \right) \psi_{-1 +\frac{1}{2}}^{\uparrow }\left(x^{\prime \prime}, \bfp^{\prime \prime} \right) + \psi_{+1 -\frac{1}{2}}^{\uparrow \ast}\left(x^\prime, \bfp^{\prime} \right) \psi_{-1 -\frac{1}{2}}^{\uparrow }\left(x^{\prime \prime}, \bfp^{\prime \prime} \right) \nonumber \\
   & +&\psi_{-1 +\frac{1}{2}}^{\uparrow \ast}\left(x^\prime, \bfp^{\prime}\right) \psi_{+1 +\frac{1}{2}}^{\uparrow }\left(x^{\prime \prime}, \bfp^{\prime \prime} \right)+\psi_{-1 -\frac{1}{2}}^{\uparrow \ast}\left(x^\prime, \bfp^{\prime} \right) \psi_{+1 -\frac{1}{2}}^{\uparrow }\left(x^{\prime \prime}, \bfp^{\prime \prime} \right) \Big] 
\end{eqnarray}
\begin{eqnarray}\label{eq:t4tilde}
    \widetilde{T}^g_{4} &= & \int \frac{d^2 \textbf{p}_\perp}{16 \pi^3} \Big[ \psi_{+1 +\frac{1}{2}}^{\uparrow \ast}\left(x^\prime, \bfp^{\prime} \right) \psi_{-1 +\frac{1}{2}}^{\uparrow }\left(x^{\prime \prime}, \bfp^{\prime \prime} \right) + \psi_{+1 -\frac{1}{2}}^{\uparrow \ast}\left(x^\prime, \bfp^{\prime} \right) \psi_{-1 -\frac{1}{2}}^{\uparrow }\left(x^{\prime \prime}, \bfp^{\prime \prime} \right) \nonumber \\
   & - &\psi_{-1 +\frac{1}{2}}^{\uparrow \ast}\left(x^\prime, \bfp^{\prime}\right) \psi_{+1 +\frac{1}{2}}^{\uparrow }\left(x^{\prime \prime}, \bfp^{\prime \prime} \right)-\psi_{-1 -\frac{1}{2}}^{\uparrow \ast}\left(x^\prime, \bfp^{\prime} \right) \psi_{+1 -\frac{1}{2}}^{\uparrow }\left(x^{\prime \prime}, \bfp^{\prime \prime} \right) \Big] 
\end{eqnarray}
where, the arguments of the LFWFs in the above equations (\ref{eq:T1g}-\ref{eq:t4tilde}) we use the notations as 
\be
x'=\frac{x-\xi}{1-\xi}, \quad\quad\quad \bfp'=\bfp+(1-x')\frac{\bf{\Delta}_{\perp}}{2},
\ee
for the active gluon longitudinal momentum fraction $(x^{\prime})$ and transverse momentum $(\mathbf{p}_{\perp}^{\prime})$ in the final proton state
and,
\be
x''=\frac{x+\xi}{1+\xi}, \quad\quad\quad \bfp''=\bfp-(1-x'')\frac{\bf{\Delta}_{\perp}}{2}.
\ee
for gluon's longitudinal momentum fraction $(x^{\prime\prime})$ and transverse momentum $(\mathbf{p}_{\perp}^{\prime\prime})$ in the initial proton state, respectively.

By employing the notion of the helicity amplitude $A_{\lambda'\mu', \lambda\mu}$ as stated in equation (\ref{helicity amp}) and using the model expression for LFWFs, one can obtain the following form of matrix elements for chiral even and odd GPDs.
\begin{eqnarray}
	T_{1}^{g}(x',x'')&=& N_{g}^{2}\Bigg[F_{2}(x',x'')  \left\{\frac{1}{\mathcal{A}^{2}}+  \left(\frac{\mathcal{B}^2}{4\mathcal{A}^{2}}-\frac{1}{4}(1-x')(1-x'')+(x''-x')\frac{\mathcal{B}}{4\mathcal{A}}\right)\frac{Q^{2}}{\mathcal{A}}\right\} \nonumber \\ 
	& & \qquad \qquad \qquad \qquad \qquad \qquad \qquad \quad -  F_{1}(x',x'')\frac{1}{\mathcal{A}} \Bigg]\exp\left[Q^{2}\left(\mathcal{C}+\frac{\mathcal{B}^{2}}{4\mathcal{A}}\right)\right],
	\label{t1g}
\end{eqnarray}
\begin{eqnarray}
	T_{2}^{g}(x',x'')&=&N_{g}^{2}\bigg[F_{3}(x',x'')\left\{\frac{1}{\mathcal{A}^{2}}+\left(\frac{\mathcal{B}^2}{4\mathcal{A}^{2}}-\frac{1}{4}(1-x')(1-x'')+(x''-x')\frac{\mathcal{B}}{4\mathcal{A}}\right)\frac{Q^{2}}{\mathcal{A}}\right\}  \nonumber \\
	& & \qquad \qquad \qquad \qquad \qquad \qquad \qquad \quad - F_{1}(x',x'')\frac{1}{\mathcal{A}}	\bigg]\exp\left[Q^{2}\left(\mathcal{C}+\frac{\mathcal{B}^{2}}{4\mathcal{A}}\right)\right],
\end{eqnarray}

\begin{eqnarray}
    T_{3}^{g}(x',x'')=N_{g}^{2}\left[F_4(x',x'')\Big\{\frac{\mathcal{B}Q}{2\mathcal{A}^2}+\frac{Q}{2\mathcal{A}}(1-x'')\Big\} - F_5(x',x'')
\Big\{\frac{\mathcal{B}Q}{2\mathcal{A}^2}-\frac{Q}{2\mathcal{A}}(1-x')\Big\}\right]
\exp\Big[Q^2\Big(\mathcal{C}+\frac{\mathcal{B}^2}{4\mathcal{A}}\Big)\Big],
\end{eqnarray}
\begin{eqnarray}
    T_{4}^{g}(x',x'')=N_{g}^{2}\left[F_4(x',x'')\Big\{\frac{\mathcal{B}Q}{2\mathcal{A}^2}+\frac{Q}{2\mathcal{A}}(1-x'')\Big\} + F_5(x',x'')
\Big\{\frac{\mathcal{B}Q}{2\mathcal{A}^2}-\frac{Q}{2\mathcal{A}}(1-x')\Big\}\right]
\exp\Big[Q^2\Big(\mathcal{C}+\frac{\mathcal{B}^2}{4\mathcal{A}}\Big)\Big] \label{T3F},
\end{eqnarray}

\begin{eqnarray}
	\widetilde{T}^g_{1}(x',x'') =N_{g}^{2}\bigg[\frac{F_5(x',x'')}{1-x'}\Big\{\frac{\mathcal{B}Q}{2\mathcal{A}^2}-\frac{Q}{2\mathcal{A}}(1-x')\Big\} 
 -\frac{F_4(x',x'')}{1-x''}
	\Big\{\frac{\mathcal{B}Q}{2\mathcal{A}^2}+\frac{Q}{2\mathcal{A}}(1-x'')\Big\}\bigg]
	\exp\Big[Q^2\Big(\mathcal{C}+\frac{\mathcal{B}^2}{4\mathcal{A}}\Big)\Big] ,
\end{eqnarray}
\be
\widetilde{T}^g_{2}(x',x'')=\widetilde{T}^g_{1}(x',x'')
\ee
\begin{eqnarray}
	\widetilde{T}^g_{3}(x',x'')&=& N_{g}^{2}\left(F_{6}(x',x'')- F_{7}(x',x'')\right) \exp\left[Q^{2}\left(\mathcal{C}+\frac{\mathcal{B}^{2}}{4\mathcal{A}}\right)\right] \nonumber \\
	&&	\times \left\{\left(\frac{\mathcal{\mathcal{B}}^2}{4\mathcal{\mathcal{A}}^{2}}-\frac{1}{4}(1-x')(1-x'')+(x''-x')\frac{\mathcal{B}}{4\mathcal{A}}\right)\frac{Q^{2}}{\mathcal{A}}\right\}
\end{eqnarray}
\begin{eqnarray}
	\widetilde{T}^g_{4}(x',x'')& =& N_{g}^{2}\left(F_{6}(x',x'')+F_{7}(x',x'')\right) \exp\left[Q^{2}\left(\mathcal{C}+\frac{\mathcal{B}^{2}}{4\mathcal{A}}\right)\right] \nonumber \\
	&&	\times \left\{\left(\frac{\mathcal{\mathcal{B}}^2}{4\mathcal{\mathcal{A}}^{2}}-\frac{1}{4}(1-x')(1-x'')+(x''-x')\frac{\mathcal{B}}{4\mathcal{A}}\right)\frac{Q^{2}}{\mathcal{A}}\right\}
	\label{t4tilde}
\end{eqnarray}

where $\mathcal{A}, \mathcal{B}$ and $\mathcal{C}$ are defined as,
\begin{eqnarray}
  \mathcal{A}&=&\mathcal{A}^{g}(x',x'')=-\frac{1}{2\kappa^{2}}\left(\frac{\log(\frac{1}{1-x'})}{x'^{2}}+\frac{\log(\frac{1}{1-x''})}{x''^{2}}\right)\nonumber\\
  \mathcal{B}&=&\mathcal{B}^{g}(x',x'')=\frac{1}{2\kappa^{2}}\left(\frac{(1-x')\log(\frac{1}{1-x'})}{x'^{2}}-\frac{(1-x'')\log(\frac{1}{1-x''})}{x''^{2}}\right)\nonumber \\
  \mathcal{C}&=&\mathcal{C}^{g}(x',x'')=\frac{1}{2\kappa^{2}}\left(\frac{(1-x')^{2}\log(\frac{1}{1-x'})}{4x'^{2}}+\frac{(1-x'')^{2}\log(\frac{1}{1-x''})}{4x''^{2}}\right)
\end{eqnarray}
whereas the functions, $F_{i}$ are para as,
\begin{eqnarray}
    F_{1}(x',x'')&=&\frac{2}{\kappa^{2}}\sqrt{\frac{\log(\frac{1}{1-x'})}{x'}\frac{\log(\frac{1}{1-x''})}{x''}}(x'x'')^{b}\left((1-x')(1-x'')\right)^{a}\left(M-\frac{M_{X}}{1-x'}\right)\left(M-\frac{M_{X}}{1-x''}\right)\nonumber\\
    F_{2}(x',x'')&=&\frac{2}{\kappa^{2}}\sqrt{\frac{\log(\frac{1}{1-x'})}{x'}\frac{\log(\frac{1}{1-x''})}{x''}}(x'x'')^{b}\left((1-x')(1-x'')\right)^{a}\left(\frac{1+(1-x')(1-x'')}{x'x''(1-x')(1-x'')}\right)\nonumber\\
    F_{3}(x',x'')&=&\frac{2}{\kappa^{2}}\sqrt{\frac{\log(\frac{1}{1-x'})}{x'}\frac{\log(\frac{1}{1-x''})}{x''}}(x'x'')^{b}\left((1-x')(1-x'')\right)^{a}\left(\frac{1-(1-x')(1-x'')}{x'x''(1-x')(1-x'')}\right) \nonumber \\
    F_{4}(x',x'')&=&\frac{2}{\kappa^{2}}\sqrt{\frac{\log(\frac{1}{1-x'})}{x'}\frac{\log(\frac{1}{1-x''})}{x''}}(x'x'')^{b}\left((1-x')(1-x'')\right)^{a}\frac{1}{x''}\left(M-\frac{M_{X}}{1-x'}\right) \nonumber\\
     F_{5}(x',x'')&=&\frac{2}{\kappa^{2}}\sqrt{\frac{\log(\frac{1}{1-x'})}{x'}\frac{\log(\frac{1}{1-x''})}{x''}}(x'x'')^{b}\left((1-x')(1-x'')\right)^{a}\frac{1}{x'}\left(M-\frac{M_{X}}{1-x''}\right) \nonumber\\
      F_{6}(x',x'')&=&\frac{2}{\kappa^{2}}\sqrt{\frac{\log(\frac{1}{1-x'})}{x'}\frac{\log(\frac{1}{1-x''})}{x''}}(x'x'')^{b}\left((1-x')(1-x'')\right)^{a}  \frac{1}{x^{\prime \prime} x^{\prime} (1-x^{\prime})} \nonumber \\
      F_{7}(x',x'')&=&\frac{2}{\kappa^{2}}\sqrt{\frac{\log(\frac{1}{1-x'})}{x'}\frac{\log(\frac{1}{1-x''})}{x''}}(x'x'')^{b}\left((1-x')(1-x'')\right)^{a}  \frac{1}{x^{\prime} x^{\prime \prime} (1-x^{\prime \prime})} 
\end{eqnarray}

\bibliography{Ref.bib}

\end{document}